\documentclass[sigconf, nonacm]{acmart}

\usepackage[]{hyperref}

\usepackage{graphicx}
\usepackage{float}
\usepackage{url}

\usepackage{multirow}
\usepackage{wasysym}
\usepackage{algorithm}
\usepackage[noend]{algorithmic}

\usepackage{makecell}
\newcommand{\etal}{\emph{et al.}}

\usepackage{pifont}

\usepackage{arydshln}
\usepackage{enumitem}

\usepackage{fancyhdr}

\usepackage{balance} 

\usepackage{bbding}

\AtBeginDocument{%
  }

\begin{document}
\begin{sloppypar}

\pagestyle{fancy}
\fancyhf{} 
\fancyfoot[C]{\thepage} 

\title{Dynamic Vulnerability Patching for Heterogeneous Embedded Systems Using Stack Frame Reconstruction}

\settopmatter{authorsperrow=4}
\author{Ming Zhou}
\orcid{0009-0005-6873-5710}
\email{mingzhou@njust.edu.cn}
\affiliation{
  \institution{SCS, Nanjing University of Science and Technology}
  \city{Nanjing}
  \country{China}}

\author{Xupu Hu}
\orcid{0009-0002-8896-1203}
\email{huxupu@njust.edu.cn}
\affiliation{
  \institution{SCS, Nanjing University of Science and Technology}
  \city{Nanjing}
  \country{China}}

\author{Zhihao Wang}
\orcid{0000-0002-0144-889X}
\email{zhihao@ustc.edu}
\affiliation{
  \institution{Purple Mountain Laboratories}
  \city{Nanjing}
  \country{China}}

\author{Haining Wang}
\orcid{0000-0002-9665-7511}
\email{hnw@vt.edu}
\affiliation{
  \institution{ECE, Virginia Tech}
  \city{Arlington, Virginia}
  \country{USA}}

\author{Hui Wen}
\orcid{0000-0002-3786-3358}
\email{wenhui@iie.ac.cn}
\affiliation{
  \institution{Institute of Information Engineering, CAS}
  \city{Beijng}
  \country{China}}

\author{Limin Sun}
\orcid{0000-0002-6578-0680}
\email{sunlimin@iie.ac.cn}
\affiliation{
  \institution{Institute of Information Engineering, CAS}
  \city{Beijng}
  \country{China}}

\author{Peng Zhang}
\orcid{0000-0001-9518-5914}
\email{zhang_peng@njust.edu.cn}
\affiliation{
  \institution{SCS, Nanjing University of Science and Technology}
  \city{Nanjing}
  \country{China}}


\renewcommand{\shortauthors}{Ming Zhou et al. }

\begin{abstract}
Existing dynamic vulnerability patching techniques are not well-suited for embedded devices, especially mission-critical ones such as medical equipment, as they have limited computational power and memory but uninterrupted service requirements. 
Those devices often lack sufficient idle memory for dynamic patching, and the diverse architectures of embedded systems further complicate the creation of patch triggers that are compatible across various system kernels and hardware platforms. 
To address these challenges, we propose a hot patching framework called StackPatch that facilitates patch development based on stack frame reconstruction. StackPatch introduces different triggering strategies to update programs stored in memory units. We leverage the exception-handling mechanisms commonly available in embedded processors to enhance StackPatch's adaptability across different processor architectures for control flow redirection. 
We evaluated StackPatch on embedded devices featuring three major microcontroller (MCU) architectures: ARM, RISC-V, and Xtensa. In the experiments, we used StackPatch to successfully fix 102 publicly disclosed vulnerabilities in real-time operating systems (RTOS). We applied patching to medical devices, soft programmable logic controllers (PLCs), and network services, with StackPatch consistently completing each vulnerability remediation in less than 260 MCU clock cycles. 
\end{abstract}

%
%

\ccsdesc[500]{Computer systems organization~Firmware}
\ccsdesc[500]{Security and privacy~Vulnerability management}

\keywords{Real-time Embedded Systems, Dynamic Security Patch, Stack Frame Reconstruction, Control Flow Redirection}

\received{14 April 2025}
\received[revised]{1 July 2025}
\received[accepted]{13 August 2025}

\maketitle

\section{Introduction}
Embedded systems are widely used in mission-critical tasks such as smart medical equipment and industrial control systems, making them particularly attractive targets for security exploitation. 
For example, Stuxnet exploited a Windows LNK vulnerability to manipulate control programs in Programmable Logic Controllers (PLCs) and alter centrifuge speeds~\cite{Falliere}. 
Similarly, adversaries intercepted communication data from medical pacemakers using reverse engineering and eavesdropping techniques, compromising victims' private information and even posing a life threat~\cite{Halperin}. 
Furthermore, a series of zero-day vulnerabilities in TCP/IP stacks have posed serious security risks to embedded systems~\cite{Ripple20}. Timely patching of vulnerabilities in embedded devices is therefore essential to ensuring the security of mission-critical systems. 

Dynamic vulnerability patching addresses program vulnerabilities within constrained time frames by using three main techniques: relocation of linked binaries, memory redundancy, and instrumentation. 
Tools such as Katana~\cite{Ramaswamy} and Kitsune~\cite{Hayden} rely on the operating system's support for dynamic linking to achieve binary relocation. However, embedded systems often rely on statically linked binaries, and many operating systems do not support dynamic linking, further complicating the use of this technique in embedded environments~\cite{Zurawski}. 
To address this issue, Holmbacka \etal~\cite{holmbacka} proposed componentizing FreeRTOS tasks into independent Executable and Linkable Format (ELF) binaries and implemented dynamic linking and relinking of FreeRTOS tasks through the addition of runtime mechanisms. 
However, this solution has limitations: it is only applicable to updates of RTOS tasks, cannot be used for updating the kernel, and does not support firmware updates for bare-metal devices such as PLCs. 

Memory redundancy, also known as the A/B scheme, maintains two system instances simultaneously, one active and the other available for update. It is used by Android devices~\cite{androidAB} and Espressif ESP32 microcontrollers~\cite{espAB} for device updates. 
However, the need for double the memory makes it impractical for resource-constrained embedded devices. In addition, this approach often requires a complete reboot of the device. 
UpStare~\cite{Makris} relies on software signals and instrumentation to trigger updates in user‑space processes. 
However, most bare metal and embedded RTOSs lack signal infrastructures and user‑level context switching, so the same technique cannot be applied in those environments. 
WfPatch~\cite{Florian} uses instrumentation to detect and modify the instructions of running programs. However, for embedded programs running in ROM, modifying the code requires erasing an entire ROM sector before writing, which introduces significant update delays. 

HERA~\cite{Niesler} utilizes the Flash Patch and Breakpoint Unit (FPB)~\cite{fpb} on ARM Cortex-M3/M4 microcontroller unit (MCU) to trigger hot patches stored in RAM, allowing firmware vulnerabilities to be patched without modifying the ROM program. However, such a solution is limited to the Cortex-M3/M4 architecture and cannot support other processor types. 
Various hardware and system configurations across different architectures (e.g., ARM, RISC-V, and Xtensa) complicate triggering vulnerability patches. 
Unlike the x86 and AMD architectures commonly used in PCs and server systems, embedded devices feature dozens of architectures such as ARM, RISC-V, Xtensa LX7, and MIPS. 
During the vulnerability patching process, device vendors need to first merge the official patch modifications into the source code, then rebuild the firmware using device-specific compilation configurations, and finally reflash the device firmware. This process is time-consuming and prone to errors~\cite{YiHe}. 
RapidPatch~\cite{YiHe} addresses this limitation by porting the eBPF virtual machine to embedded systems, enabling eBPF patches to work across heterogeneous devices. However, incorporating eBPF components increases storage and runtime overhead in embedded devices. Moreover, both HERA and RapidPatch focus on patching programs stored in Flash memory, while overlooking vulnerabilities in programs running in RAM. 
The kpatch and ksplice use the shadow variable to add fields to existing data structures without altering the original data layout. However, this technique cannot handle scenarios where the size of global variables changes~\cite{Zhenyu}. 

In this paper, we propose StackPatch, a stack frame reconstruction-based hot patching framework to address security vulnerabilities in embedded systems. Unlike traditional firmware update methods~\cite{freertosOTA, zephyrota, Dong, Panta}, which require updating entire or partial firmware images, StackPatch leverages stack frames to generate dynamic hot patches. Our approach eliminates the need for full-scale updates by constructing a mapping between vulnerable function variables and the stack pointer. 
We further extend StackPatch to support the repair of global variables and macro definitions. In cases that involve global variables, we introduce four distinct repair strategies based on memory space modifications. For patch tasks related to macro definition changes, we define update points at the locations where macros are expanded and use our stack frame mapping to translate the new macro code into a hot patch.
The core of StackPatch is its ability to harness the processor's built-in exception-handling mechanism. This mechanism allows the system to redirect the control flow from vulnerable code to the hot patch, thus minimizing runtime overhead. Unlike other systems, such as FPB or eBPF VM, StackPatch does not require additional hardware or software support, ensuring its lightness and efficiency. 

We evaluated StackPatch's performance and effectiveness across a variety of heterogeneous embedded systems, including ARM, RISC-V, and Xtensa MCUs. 
Experimental measurements show that StackPatch adds fewer than 260 MCU clock cycles of latency, well within the real‑time bounds mandated by typical RTOS deployments. 
We successfully patched 102 RTOS vulnerabilities and demonstrated the framework's applicability to critical use cases, including medical devices and PLCs, highlighting its high availability, efficiency, and broad applicability. 

The main contributions of this work are summarized as follows: 
\begin{itemize} 
\item We analyze the stack frame structures of embedded MCU architectures, including ARM, RISC-V, and Xtensa. We then develop a stack-frame reconstruction mechanism to generate hot patches. Additionally, we extend the patching capabilities to support changes to global variables and macro definitions. 
\item We modify the exception handlers to redirect the control for adaptation to heterogeneous embedded systems. StackPatch enables the selection of appropriate hot patch triggering strategies based on a program's storage location. 
\item We apply StackPatch to medical devices, soft PLCs, and network services, patching 102 vulnerabilities across four embedded systems and three MCU architectures. We also compare StackPatch with four advanced patching approaches. 
\end{itemize}

The remainder of this paper is structured as follows. 
Section~\ref{section:background} introduces dynamic patching. 
Section~\ref{section:stackPatch_design} details the design of StackPatch and how it leverages stack frames for vulnerability patching. 
Section~\ref{section:evaluation} presents a comprehensive evaluation of StackPatch. 
Section~\ref{section:related_work} surveys related work. 
Section~\ref{section:discussion} discusses some limitations of StackPatch, and finally, Section~\ref{section:conclusion} concludes the paper.

\section{Dynamic Patching}
\label{section:background}
In this section, we introduce dynamic patching in embedded systems, outline the characteristics that make embedded systems suitable for dynamic patching, describe common strategies for triggering patches, and discuss challenges in patching heterogeneous environments. 

\subsection{Dynamic Patching in Embedded Systems}
\label{background:dynamic_security_patch}
Given ($\Pi$, $s$), where $\Pi$ is the program code and $s$ is an execution state. We define $s$ = ($md, gr, pc$), where $md$ is the current processor mode (e.g., user or kernel mode), $gr$ is the set of general registers, and $pc$ represents the current value of the program counter (PC). 
Dynamic patching in embedded systems can be reduced to a state transition problem: constructing a transition mechanism $F$ such that $s'$ = $F$ ($s$), where $s$ represents the execution state of the vulnerable code $\Pi$ and $s'$ represents the execution state of the benign code $\Pi'$. 

Existing approaches use various transition mechanisms $F$. In the A/B scheme, $F$ replaces the entire vulnerable code with the patched code. Other approaches, such as HERA~\cite{Niesler}, RapidPatch~\cite{YiHe}, and RLPatch\cite{MingZhou}, rely on code reconstruction. 
However, these approaches often fail to account for differences in MCU architecture and exception handling when saving and restoring execution states, leading to high overhead. 
To address these limitations, we introduce a stack-frame reconstruction approach to implement the transition mechanism $F$. 
This approach involves creating an initial stack frame $f (gr,ra)$, reconstructing it to $f' (gr,ra)$, and finally restoring it to $f' (gr,ra)$. 
Using the exception handler, we create $f$ to save the execution state $s$ before the vulnerable code executes. The stack frame $f$ is then modified within the patch code. Lastly, we restore the stack frame $f'$ using the exception handler to execute the benign code $\Pi'$ in the new state $s'$. Here, $f (gr, ra)$ refers to the stack frame, where $gr$ represents the general registers, and $ra$ is the return address. 

In addition to satisfying $F$, dynamic patching in real-time embedded systems must meet five key conditions (Table~\ref{tab:methods}): support system without reboot (\textbf{T1}), support for heterogeneous systems (\textbf{T2}), support for multiple triggering mechanisms (\textbf{T3}), lightweight deployment (\textbf{T4}), and support for high-complexity patch (\textbf{T5}). Note that \textbf{T5} requires system support for complex patch types and low-complexity patch generation. However, existing approaches do not meet all these requirements simultaneously. 

As shown in Table~\ref{tab:methods}, the A/B scheme requires a system reboot to apply the new program version (violating \textbf{T1} and \textbf{T3}) and doubles the memory needed to store both program versions (violating \textbf{T4}). 
HERA relies on the FPB-based patch triggering mechanism, available only on ARM Cortex-M3/M4 architectures. It requires patch deployment in assembly language, demanding advanced programming skills, and only supports official patches that involve simple instruction modifications (violating \textbf{T2}, \textbf{T3}, and \textbf{T5}). 
RapidPatch uses hardware breakpoints and hooks as the patch trigger mechanism (partially satisfying \textbf{T3}). However, its use of eBPF introduces significant memory overhead and requires converting C programs into eBPF bytecode, complicating the patch development process (violating \textbf{T4}). In addition, it requires converting C programs into eBPF bytecode, which complicates the patch development process (violating \textbf{T5}). 
AutoPatch only supports instrumentation-based patch triggering (violating \textbf{T3}). It requires setting numerous instrumentation locations before compilation, which leads to significant computational overhead due to frequent control flow jumps (violating \textbf{T4}). 
While AutoPatch uses the weakest precondition reasoning to generate hot patches, it does not support patching changes to data structures or macro definitions (violating \textbf{T5}). 
Lastly, RLPatch supports trigger strategies using both hardware and software breakpoints (partially satisfying \textbf{T3}). However, it does not account for heterogeneous MCU architectures (violating \textbf{T2}). RLPatch requires patching in assembly language and only supports simple patches (violating \textbf{T5}). 

\begin{table}[t]
    \centering
    \normalsize
    \begin{tabular}{l|c|c|c|c|c}
    \hline
        ~ & \textbf{T1} & \textbf{T2} & \textbf{T3} & \textbf{T4} &\textbf{T5} \\ \hline
        A/B scheme& $\CIRCLE$ & $\Circle$&$\CIRCLE$  & $\CIRCLE$ &$\Circle$ \\ \hline
        HERA~\cite{Niesler} & $\Circle$ & $\CIRCLE$&$\CIRCLE$& $\Circle$ &$\CIRCLE$ \\ \hline
        RapidPatch~\cite{YiHe} & $\Circle$ & $\Circle$&$\RIGHTcircle$& $\CIRCLE$ &$\CIRCLE$ \\ \hline
        AutoPatch~\cite{Salehi}& $\Circle$ & $\Circle$&$\CIRCLE$& $\CIRCLE$ &$\CIRCLE$ \\ \hline
        RLPatch~\cite{MingZhou} & $\Circle$ & $\CIRCLE$&$\RIGHTcircle$& $\Circle$ &$\CIRCLE$ \\ \hline
    \end{tabular}
    \caption{A summary of how existing dynamic patching approaches for real-time embedded systems meet various requirements. \textbf{T1}: Support system without reboot. \textbf{T2}: Support multiple processor architectures. \textbf{T3}: Support multiple triggering mechanisms. \textbf{T4}: Lightweight solution. \textbf{T5}: Support high-complexity patches. The $\Circle$ indicates the technique is satisfied, $\RIGHTcircle$ indicates the technique is partially satisfied, and $\CIRCLE$ indicates the technique is not satisfied.}
    \label{tab:methods}
\end{table}

\begin{table*}[t]
    \centering
     \normalsize
    \begin{tabular}{l|l|l|l|l}
        \hline
        \multicolumn{2}{c|}{\textbf{Patch triggering mechanism}} & \textbf{MCU architecture} & \textbf{Memory type} & \textbf{Num. of patches} \\ \hline
        \multirow{2}*{Hardware-based} & hardware breakpoints & ARM Cortex-M3$\sim$M55, RISC-V & RAM \& Flash & 2$\sim$8 \\ \cline{2-5} 
        \multicolumn{1}{c|}{} & FPB with remapping & ARM Cortex M3/M4 & Flash & 6 \\ \hline
        \multirow{2}*{Software-based} & software breakpoints & ARM, RISC-V, Xtensa & RAM & unlimited \\ \cline{2-5} 
        \multicolumn{1}{c|}{} & hook & ARM, RISC-V, Xtensa & RAM \& Flash & unlimited \\ 
        \hline
    \end{tabular}
    \caption{Dynamic patch triggering strategies in real-time embedded devices.}
    \label{tab:table1}
\end{table*}

\subsection{Embedded Systems Characteristics}
\label{background:realtime_embedded_system}
\textbf{Embedded devices} operate under tight storage and computing budgets, so they typically adopt low‑power MCUs. 
To guarantee predictable timing, these MCUs run lightweight real‑time operating systems (RTOSs) such as FreeRTOS~\cite{freertos} or Zephyr~\cite{zephyr}, which enforce strict deadline guarantees. 
Real‑time workloads are usually classified as hard, firm, or soft, based on the impact of missing a deadline. 
In hard real‑time settings, such as medical monitors (e.g., heart‑rate sensors) and industrial controllers (e.g., PLCs), every deadline must be met; even a single overrun can compromise safety or cause intolerable system failure. 

To ensure reliability and robustness, embedded systems use \textbf{exception-handling mechanisms} to detect and recover from errors. These mechanisms manage exceptions using interrupts, allowing developers to define specific triggers based on the type of exception. For instance, in ARM architectures, breakpoints trigger the \texttt{DebugMon\_Handler} in debug monitor mode, while undefined instructions trigger the \texttt{Hardfault\_Handler}. 
We found that the built-in exception-handling mechanisms in embedded devices can redirect control flow to patch code without causing downtime, thereby satisfying \textbf{T1} (no reboot), \textbf{T2} (support for heterogeneous systems), and \textbf{T4} (lightweight deployment). 
Most embedded architectures lack a Memory Management Unit (MMU), so they cannot provide memory isolation mechanisms, allowing user-space and kernel-space data to be directly shared~\cite{YiHe}. In this case, exception handlers can modify any memory without restrictions, including code in both user-space and kernel-space. 

A \textbf{stack frame} is a dedicated memory region used during function calls. When an MCU enters an exception, the stack frame stores the values of general-purpose registers and restores them upon exiting the exception. Stack frame operations vary depending on the MCU's instruction set, register usage, and calling conventions. For instance, in ARM architectures, some general-purpose registers are automatically saved to the stack frame upon entering the exception handler, whereas in RISC-V and Xtensa architectures, the register values must be saved manually using assembly code. 
We discovered that mapping the data in the stack frame to C language pointers can greatly simplify the patch code development process, addressing \textbf{T5} (ease of patch development). 
Specifically, we allocate a stack frame $f$ and uniformly save the general-purpose register values $gr$ and return address $ra$ in the exception handler before the vulnerable code executes. During this patch development, we modify the saved contents in the stack frame, enabling us to patch the vulnerable program effectively.

\subsection{Patch Triggering Strategies}
\label{background:trittering_strategies}
Table~\ref{tab:table1} summarizes two dynamic patch-triggering strategies in real-time embedded devices: hardware-based and software-based triggers. 
Hardware-based triggers store the breakpoint address (the entry point of the vulnerability code) in dedicated registers, making this approach highly efficient. However, the number of vulnerabilities that can be patched is limited by the number of available registers, typically no more than eight. Hardware-based triggers operate by comparing the value in the hardware register with the Program Counter (PC) to throw exceptions. 
This includes hardware breakpoints and the Flash Patch and Breakpoint (FPB) unit. Hardware breakpoints can be applied to programs stored in any memory, including ROM. 
The FPB unit supports address remapping and therefore offers a fundamental hardware‑level debugging mechanism. HERA leverages this remapping capability to trigger updates.

Software triggers are typically implemented with software breakpoints or hooks. A software breakpoint\textemdash supported by almost every MCU architecture\textemdash raises an exception by replacing the original instruction with an invalid opcode, enabling precise control over execution flow. 
Examples include the \texttt{bkpt} instruction on ARM, the \texttt{ebreak} instruction on RISC-V, and the \texttt{break} instruction on Xtensa. 
During compilation, hooks place patch triggers at designated program points, such as the entry of a vulnerable function. 
Each trigger serves as a trampoline: it overwrites the original instruction with a branch or jump that forwards execution to the patch routine. 
Given sufficient memory, software triggers can cover an unlimited number of vulnerabilities, although they incur a modest runtime overhead. 

Different storage types require specific triggering mechanisms for patching. 
The FPB unit can only patch programs stored in Flash memory using its hardware breakpoint functionality. The Data Watchpoint and Trace (DWT) on ARM architecture can patch both Flash and RAM programs. 
However, FPB-based hardware breakpoints are limited to Flash programs, while software breakpoints are effective for patching programs stored in RAM. Hooks can be used to patch programs in both RAM and Flash. 
To support heterogeneous dynamic security patching, StackPatch implements all three triggering mechanisms: hardware breakpoints, software breakpoints, and hooks.

\subsection{Heterogeneous Environment Patching}
\label{background:heterogeneous environment}
Real-time operating system (RTOS) vendors typically release new versions to address security vulnerabilities. However, there is often a significant delay before these updates reach end devices.
For example, systems like FreeRTOS and Zephyr typically require 3 to 6 months to complete an upgrade, while TizenRT may take over a year to patch known vulnerabilities~\cite{YiHe}. 
During this time, effective exploit strategies for specific vulnerabilities may become publicly available, sometimes before or immediately after the release of patches. 
RTOS platforms mainly rely on Over-the-Air (OTA) technology to update entire firmware images~\cite{freertosOTA, zephyrota}, enabling manufacturers to remotely push new firmware to devices without requiring physical connections. 
To reduce delays in patching, systems like R3 and Zephyr employ incremental reprogramming techniques that minimize the amount of data transmitted wirelessly~\cite{Dong, Panta}. 
However, these approaches often interrupt services running on embedded devices, leaving real-time constrained systems vulnerable for extended periods. 

The need to support multiple processor architectures in StackPatch arises from the growing diversity of embedded systems in modern applications. 
As embedded systems become increasingly specialized\textemdash such as in medical devices and industrial control systems\textemdash compatibility across different MCU architectures becomes essential, particularly in heterogeneous environments where multiple processor types may coexist. 
StackPatch addresses this challenge by providing a flexible patching mechanism that can be adapted to various MCU architectures, ensuring broad applicability across different real-time systems. 
This approach enables StackPatch to deliver a unified solution for multiple platforms while maintaining efficiency and minimizing overhead.

\section{StackPatch Design}
\label{section:stackPatch_design}

\begin{figure*}[t]
    \centering
    \includegraphics[width=\linewidth]{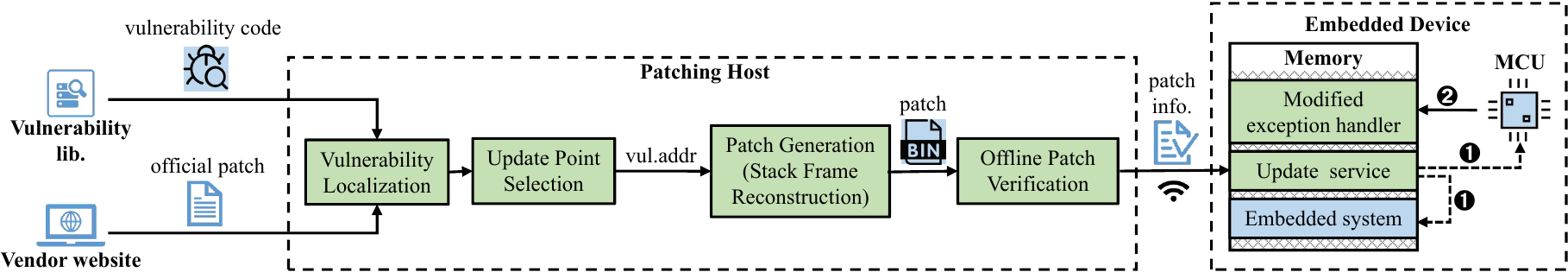}
    \caption{Overview of StackPatch.}
    \Description{Overview of StackPatch}
    \label{fig:Overview}
\end{figure*}

\subsection{Overview}
\label{stackPatch:overview}
Figure~\ref{fig:Overview} shows an overview of StackPatch. On the patching host side, StackPatch retrieves the vulnerability code and its descriptions from a public vulnerability database, along with the official patch code and device-specific details from the manufacturer's website. 
Its vulnerability localization component identifies the range of vulnerable code, while its update point detection component locates the vulnerability's entry point, designating it as the update point for repair. 
Once this location is determined, StackPatch generates hot patches through stack frame reconstruction and filters the unsafe patches. 
On the embedded device side, the update service establishes a connection with the patch host via wireless or wired channels to receive patch information. It then sets the update points using either hardware-based (\ding{182}) or software-based (\ding{183}) strategies. 

\begin{figure}[t]
    \centering
    \includegraphics[width=0.47\textwidth]{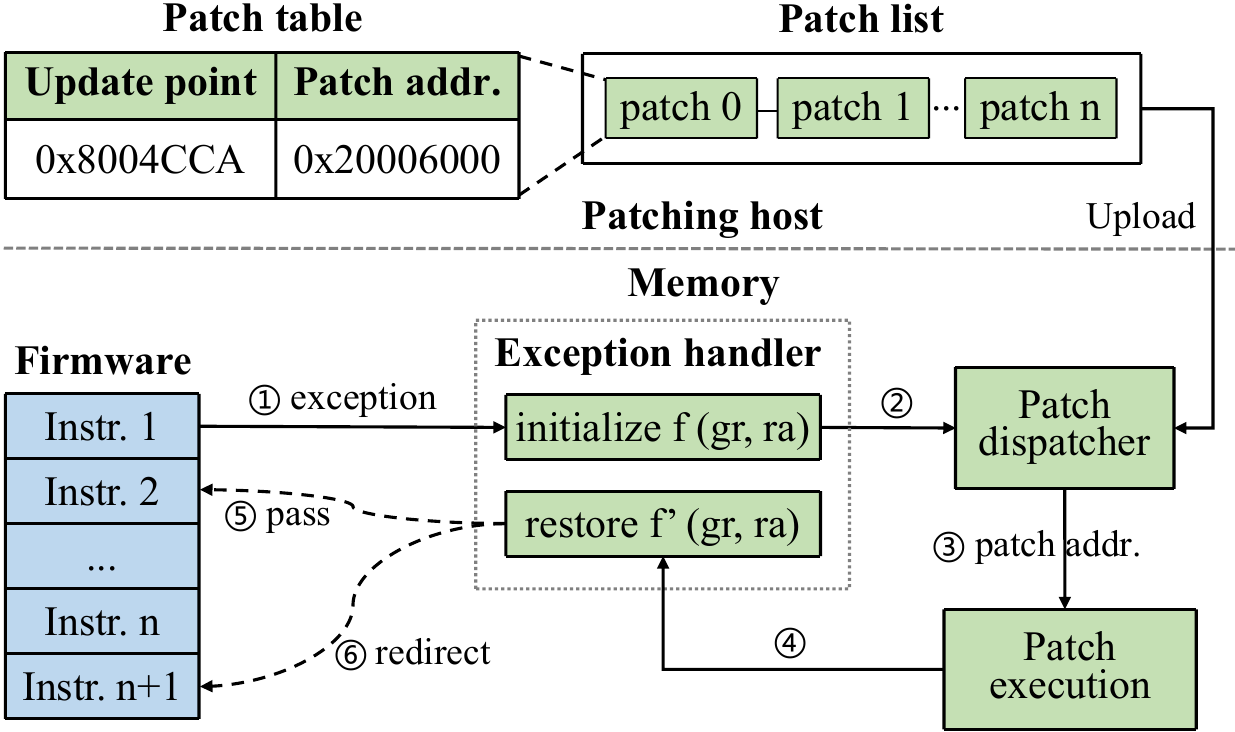}
    \caption{Execution flow of StackPatch.}
     \Description{Execution flow of StackPatch.}
    \label{fig:Contorlflow}
\end{figure}

Once the update points are configured, StackPatch is executed. Figure~\ref{fig:Contorlflow} shows the execution flow of StackPatch. When the program counter (PC) reaches the update point of the vulnerable program $\Pi$ (in execution state $s$), the MCU triggers an exception and executes its exception handler from the interrupt vector table (\ding{192}). 
After establishing the stack frame $f (gr, ra)$, the exception handler transfers control to the patch dispatcher (\ding{193}). 
The dispatcher reads the vulnerability entry address from the stack frame, uses a binary search algorithm to locate the corresponding patch address in the patch table, and jumps to execute the patch (\ding{194}). 
Once the patch code is executed, control returns to the exception handler (\ding{195}). 

The exception handler then restores the stack frame $f' (gr, ra)$ and chooses either the pass operation (\ding{196}) or the redirect operation (\ding{197}), depending on the official patch, to transfer control to the benign code $\Pi$ in the updated execution state $s'$. 
StackPatch performs the pass or redirect operation by modifying the return address $ra$ in the stack frame $f' (gr, ra)$ (Section~\ref{stackPatch:patch_generation}). 
When several vulnerabilities must be patched\textemdash for example, if the vulnerable code is inlined at multiple locations, or a vulnerability involves changes to a macro definition that expands in several places\textemdash StackPatch creates a stack frame only when the MCU reaches the relevant update point, rather than pre‑allocating frames for every site. This on‑demand strategy eliminates unnecessary stack overhead.

\begin{figure*}[t]
    \centering
    \includegraphics[width=\textwidth]{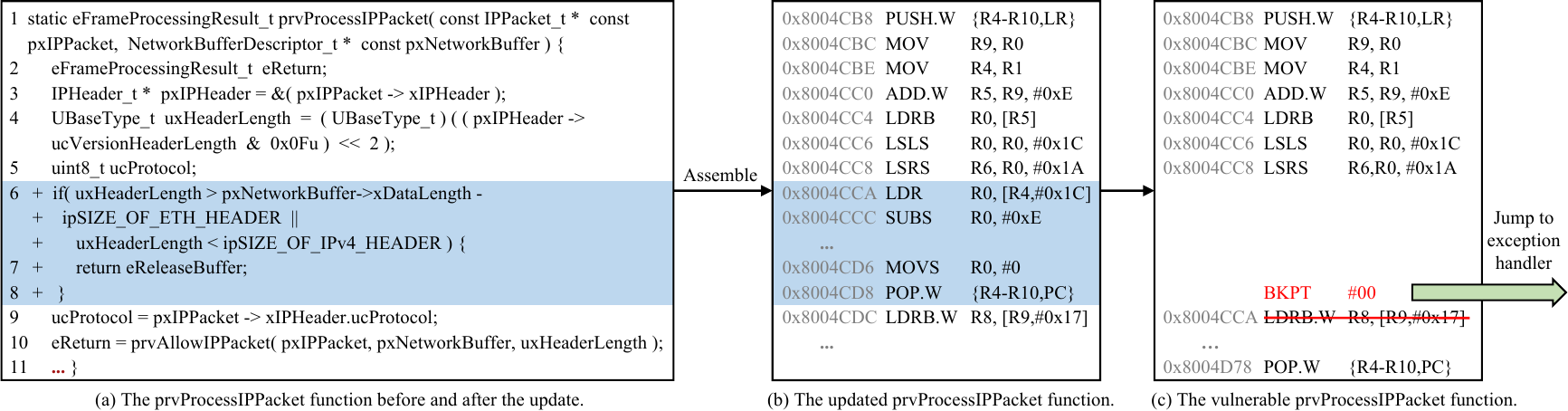}
    \caption{The process of vulnerability localization and update point selection. The code marked in blue represents the added boundary-checking functionality. The "BKPT \#00" instruction marked in red indicates the insertion of a replacement interrupt instruction at update point 0x8004CCA.}
    \Description{The process of vulnerability localization and update point selection.}
    \label{fig:2} 
\end{figure*}

\subsection{Vulnerability Localization \& Update Point Selection}
\label{stackpatch:vulnerability_detection}
The accuracy of automatic vulnerability localization is low~\cite{Yang2024LLMfTFL}, so we manually locate vulnerabilities in the firmware. 
We gather vulnerability information for real-time embedded systems from publicly available CVE databases~\cite{cveWebsite}, extracting details such as system versions, vulnerability types, MCU architectures, and triggering conditions. 
Figure~\ref{fig:2} illustrates the process of vulnerability localization and update point selection. 
Based on this data, we collect both the vulnerable code and its official patch from the Internet. 
We then disassemble both the vulnerable and updated code into assembly, tailored to the MCU architecture, to facilitate stack-frame reconstruction. 
Finally, we use the binary diffing tool BinDiff~\cite{Bindiff} to identify the differences between the official and vulnerable code, with the differences in the assembly code indicating the start and end positions of the vulnerable code. 

We use CVE-2018-16601, the case studied by HERA, to illustrate the localization of vulnerable code. This vulnerability, caused by missing boundary checks for the IP header length~\cite{CVE201816601}, affects FreeRTOS V10.0.1 and earlier, but was later fixed in subsequent releases. 
By comparing the source code of FreeRTOS V10.0.3 with that of V10.0.1, we identified that the new version adds boundary checks in the \texttt{prvProcessIPPacket} function within the \texttt{FreeRTOS\_IP.c} file (the lines marked with "+" in Figure~\ref{fig:2} (a). 
Figure~\ref{fig:2} (b) shows the updated \texttt{prvProcessIPPacket} function code after the patch was applied. 

The \textbf{update point} is the entry to the vulnerable code, and it is set at the first instance of a semantic code change between the two versions of the MCU instructions. As shown in Figure~\ref{fig:2} (c), we chose \texttt{0x8004CCA} as a safe update point, replacing the vulnerable code with a software interrupt instruction. 
If the update point is not located at the entry of the vulnerable code, the fix may be missed, resulting in a failed patch. For instance, if an address after \texttt{0x8004CCA} in Figure~\ref{fig:2} (c) is chosen as the update point, the fix fails because the boundary check is omitted. 
The update point serves as the transition point where the control flow shifts from the vulnerable code to the patch code. To minimize the complexity of patch development, the update point is selected as close to the patch code as possible. Moreover, when choosing the update point, it is critical to ensure that all variables used in the patch code have correct values. 
We use data flow analysis techniques~\cite{Ghidra} to ensure that, before execution reaches each update point, all variables referenced by the patch are properly initialized. 
Specifically, we use Ghidra's variable‑tracking feature to determine, at each instruction address, which registers currently hold the live values of local variables. 
For example, in Figure~\ref{fig:2} (a), the patch references the variable \texttt{uxHeaderLength} (line 6). 
In Figure~\ref{fig:2} (c), the value of this variable is assigned at address \texttt{0x8004CC8} and stored in the \texttt{R6} register. 
If the update point were set before \texttt{0x8004CC8}, the patch would access the incorrect value of \texttt{R6}. 
Although boundary checks can be implemented by recalculating the variable \texttt{xHeaderLength} in the hot patch, this approach increases the patch's code size and its runtime memory overhead.

\begin{figure}[t]
    \centering
    \includegraphics[width=0.475\textwidth]{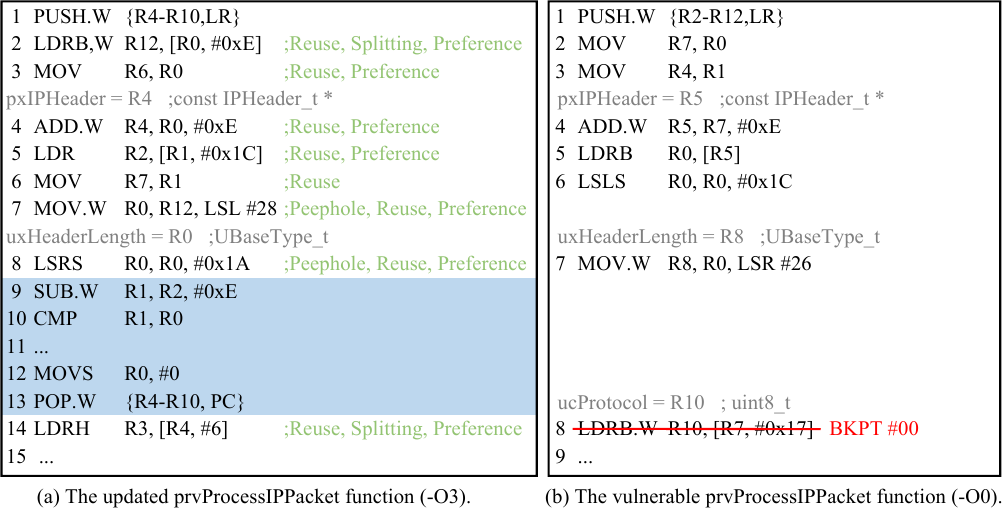}
    \caption{Binary differencing of the patched code at the O3 optimization level. Gray annotations denote variable-to-register mappings automatically recovered by Ghidra, and green annotations highlight compiler optimizations responsible for instruction-level modifications.}
    \Description{Binary differencing under high optimization levels.}
    \label{fig:diff_O3}
\end{figure}

At high optimization levels, the main obstacles to vulnerability localization are register-allocation optimizations (reuse, splitting, and preference) and peephole optimizations. 
These transformations often yield semantically equivalent code that differs substantially in opcode and register usage between versions. 
To overcome this issue, we use Ghidra's P-Code intermediate representation to automatically recover variable-to-register mappings, which mitigates structural differences introduced by optimization and enables the matching of semantically equivalent instruction sequences across patched and vulnerable binaries. 

Our procedure first consults the official source patch to identify the function entry point and the intended patch semantics. It then performs a binary-level comparison. 
Even at O3, where most instructions in the patched version have been transformed (Figure~\ref{fig:diff_O3}.a), the early variable assignments remain equivalent: the assignment to \texttt{pxIPHeader} appears at line 4 in both Figure~\ref{fig:diff_O3} (a) and Figure~\ref{fig:diff_O3} (b), and the computation of \texttt{uxHeaderLength} occurs at line 7 in Figure~\ref{fig:diff_O3} (a) and line 6 in Figure~\ref{fig:diff_O3} (b). 
The updated version in Figure~\ref{fig:diff_O3} (a) introduces the patch logic at line 9, including a bound check on \texttt{uxHeaderLength}. The vulnerable version in Figure~\ref{fig:diff_O3} (b) instead proceeds at line 8 with \texttt{LDRB.W R10, [R7, \#0x17]}, which loads \texttt{ucProtocol} without validation. 
Line 8 in Figure~\ref{fig:diff_O3} (b) is therefore the first point of semantic divergence and the preferred update point for inserting the patch. 

Update points, whether based on software or hardware breakpoints, target MCU instructions. A principle for choosing update points is that the update point cannot be set at illegal addresses or stack operations. 
Illegal addresses may cause the update point to fail, such as trying to set a software breakpoint in read-only regions like Flash, which would require erasing an entire sector. 
Setting update points on stack operations (\texttt{PUSH/POP}) can corrupt the stack frame structure, as the patch code may access uninitialized or corrupted variables. For example, if the update point is set before the operation \texttt{PUSH} (0x8004CB8) in the function \texttt{prvProcessIPPacket} (Figure~\ref{fig:2}.c), the patch will fail to retrieve the correct value of \texttt{uxHeaderLength}. 
The update point cannot be set within the update service itself, which may lead to abnormal behavior of the update service.
We replaced the first MCU instruction for the vulnerable code with a software breakpoint to set the update point at the vulnerability entry. Note that hardware breakpoints are configured directly in MCU registers. 
For hook-based solutions, the target is a function block, allowing the update point to be placed at any position in the program, providing more flexibility than breakpoint-based solutions. 

\begin{table}[t]
    \centering
    \footnotesize
      \setlength{\tabcolsep}{1.1mm}{
    \begin{tabular}{c|c|c|c|c}
        \hline
        \textbf{ R} &\textbf{* (sp + 3u)} & \makecell[c]{\textbf{* (uint32\_t *) (*}  \\ \textbf{ (sp + 1u) + 0x1C)}} & \textbf{* (sp + 9u)} & \textbf{* (sp + 15u)} \\\hline
        \textbf{uxHeaderLength} & $\times$ & ~ & ~ &\\ 
        \hline
        \makecell[c]{\textbf{pxNetworkBuffer}  \\ \textbf{->xDataLength}} & ~ &  $\times$ & ~ &\\ \hline
        \textbf{return} \textbf{value}  & ~ & ~ & $\times$& \\ \hline
        \makecell[c]{\textbf{return} \textbf{address}}  & ~ & ~ & &$\times$ \\ \hline
    \end{tabular}}
    \caption{An example of "variable-to-stack" mapping table.}
    \label{tab:mapping}
\end{table}

\begin{figure*}[t]
    \centering
    \includegraphics[width=\textwidth]{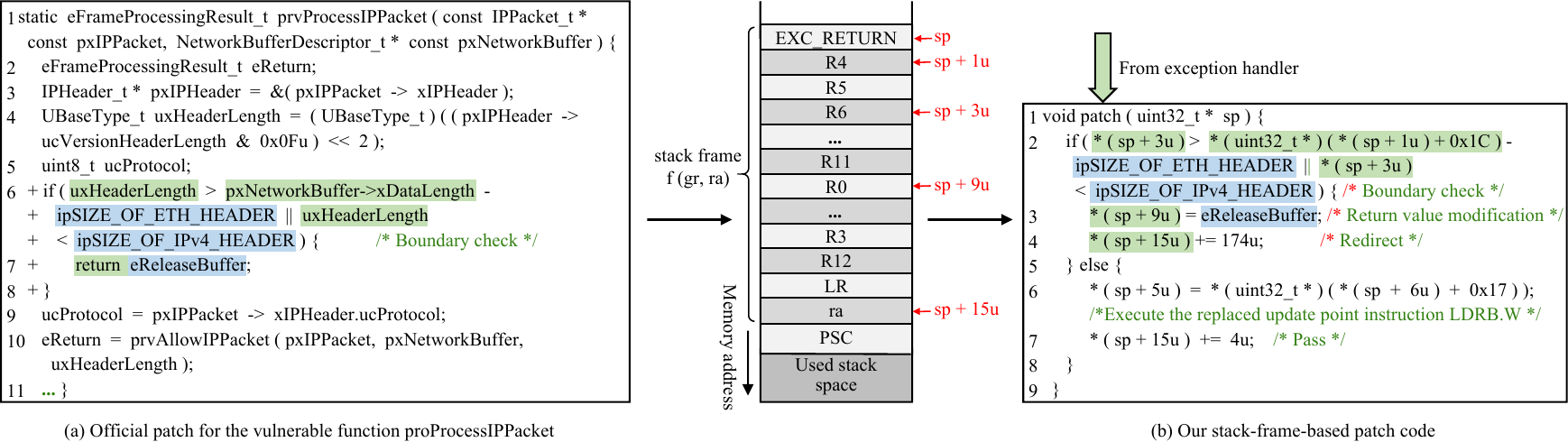}
    \caption{Stack-frame-based hot patch generation process. The stack of three MCU architectures grows downloads.}
    \Description{hot patch generation process.}
    \label{fig:3}
\end{figure*}

\subsection{Stack-Frame-Based Hot Patch Generation}
\label{stackPatch:patch_generation}
Based on the stack frame $f (gr, ra)$ established for the execution state $s$ at the update point, StackPatch constructs a mapping between \textbf{local variables} in the vulnerable function and the stack pointer using the following three steps. 
StackPatch begins with a static data flow analysis that, for each instruction address, records which register currently holds the live value of every variable. The result is a variable‑to‑register mapping $R1 = map (variable, register)$. 
Next, StackPatch establishes a register-to-stack mapping $R2 = map (register, sp\_offset)$, signifying that each register's value is stored at a particular offset ($sp\_offset$) in the stack frame. 
This mapping is derived by analyzing instruction-level operations, such as \texttt{push reg} and \texttt{str reg, [sp, \#offset]}, which are encoded in the modified exception handler (see Section~\ref{stackPatch:control_flow_redirection}). 
By composing the two mappings transitively, StackPatch derives $R = R1 \times R2$, which maps each variable to its stack-pointer offset R: $\texttt{variable} \mapsto \texttt{sp\_offset}$. 
This process is robust to compiler optimizations. 
The mapping $R1$ is built from an intermediate representation that hides machine-specific register reuse and instruction reordering, and $R2$ follows ABI-defined save/restore sequences that remain structurally stable. 
Consequently, the combined variable-to-stack mapping remains accurate across optimization levels O0, O1, O2, and O3. 

After these steps, StackPatch generates a mapping table that details the relationships between function variables and their respective stack pointer offsets in the stack frame. 
StackPatch references this mapping table to replace all function variables in the official patch with the corresponding stack pointer offsets. 
For instance, Table~\ref{tab:mapping} shows an example of a "variable-to-stack" mapping table generated from patching the function \texttt{prvProcessIPPacket}. The variable \texttt{uxHeaderLength} is replaced by \texttt{*(sp+3u)}. Additionally, \texttt{0x1C} represents the offset of \texttt{xDataLength} relative to \texttt{pxNetworkBuffer}, while macro variables such as \texttt{eReleaseBuffer}, \texttt{ipSIZE\_OF\_ETH\_HEADER} and \texttt{ipSIZE\_OF\_IPV4\_HEADER} are not involved in the mapping process. 

The mapping table also includes the relationship between the return value of the function, the return address of the function, and the stack pointer offset. 
As shown in Figure~\ref{fig:3} (a), the state \texttt{return eReleaseBuffer} consists of two operations: returning \texttt{eReleaseBuffer} as the return value (\texttt{MOVS R0, eReleaseBuffer}) and transferring control to the caller of the vulnerable function (\texttt{POP.W {R4-R10, PC}}). 
During stack frame reconstruction, the return value is mapped to \texttt{*(sp+9u)}, and the return address to \texttt{*(sp+15u)}. The 9u and 15u represent the offset of the return value and return address relative to the top of the stack frame, respectively. 
StackPatch excludes \texttt{POP} operations of official patches, as the patch code cannot directly modify the stack structure. 

Control transfer (from the exception handler to benign code) is achieved by modifying the stack frame's return address (\texttt{ra}). StackPatch supports two control transfer strategies: Redirect and Pass. 
The Redirect strategy directs the flow back to either the caller of the vulnerable function or the vulnerable function itself. 
In the first case, \texttt{ra} is modified to the vulnerable function's return instruction. 
For example, as seen in line 4 in Figure~\ref{fig:3} (b), \texttt{ra} is adjusted by $174u$ (\texttt{0x8004D78-0x8004CCA}) past the update point address. 
In the second case, \texttt{ra} points to the instruction right after the vulnerable code. The length of the vulnerable code is computed by performing a binary comparison between the updated and vulnerable code. Then \textbf{ra} is computed as the sum of the update-point address and the length of the vulnerable code. 
The Pass strategy sets \texttt{ra} to the instruction directly after the update point, with an offset matching the update point instruction's length. 
For example, ARM architecture includes instructions for both 16-bit Thumb and 32-bit Thumb-2. In one instance, the update point instruction \texttt{LDRB.W R8, [R9, \#0x17]} occupies four bytes, prompting StackPatch to offset \texttt{ra} by 4 bytes (as shown in Figure~\ref{fig:3}.b, line 7). 
Note that the choice of return strategy for the hot patch is determined by the official patch, not by the type of vulnerability. 

\begin{figure}[t]
    \centering
    \includegraphics[width=0.47\textwidth]{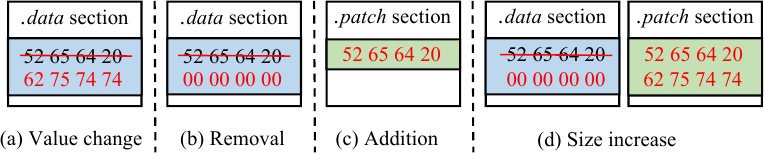}
    \caption{Memory space modifications for four global variable changes. (a) changes to variable values, (b) removed variables, (c) added variables, and (d) variable size increased.}
    \Description{globle.}
    \label{fig:global}
\end{figure}

If the official patch includes changes to \textbf{global variables}, StackPatch modifies three sections of memory: \texttt{.data}, \texttt{.patch}, and \texttt{.text}. The \texttt{.data} section stores global variable values, \texttt{.patch} section contains StackPatch's patch information, and the \texttt{.text} section holds the vulnerable program. 
We categorize changes to global variables into four types: value changes, removals, additions, and size increases,  each requiring specific memory modification strategies. 
If the official patch alters a global variable's value, StackPatch directly updates the variable in the \texttt{.data} section (Figure~\ref{fig:global}.a). 
If a global variable is removed, StackPatch sets its value to 0 in the \texttt{.data} section (Figure~\ref{fig:global}.b) and removes its reference in the \texttt{.text} section. 
For a new global variable, StackPatch adds the variable to the \texttt{.patch} section (Figure~\ref{fig:global}.c) and includes a reference in the \texttt{.text} section. 
If the size of a global variable increases, StackPatch does not modify the value in place in the \texttt{.data} section to prevent overwriting subsequent variables. Instead, it places the new value in the \texttt{.patch} section, sets the original variable's value to 0 in \texttt{.data} section (Figure~\ref{fig:global}.d), and updates its reference in the \texttt{.text} section. 
Note that the \texttt{.text} section may reside in Flash memory. To avoid long delays caused by erasing the entire vulnerability sector, the removal operation is handled by the patch program in the \texttt{.patch} section. 

If the official patch modifies a \textbf{macro variable}, StackPatch extracts the macro definitions from the firmware using the symbol table and sets the update points at the locations where the macro is expanded in the \texttt{.text} section. Then, it converts the new macro code into a hot patch using our stack frame mapping. StackPatch does not modify the macro definitions directly because these definitions are expanded during preprocessing and stored as code in the \texttt{.text} section. At runtime, only the expanded code snippets are present, making them the only modifiable targets~\cite{Kpatch}. 

After generating the C patch, we use a cross-compilation tool to produce a binary patch, during which global macros are replaced with their specific values. StackPatch then appends a trampoline instruction at the end of the patch to return control flow to the exception handler. 

\subsection{Patch Verification \& Update Service}
\label{stackPatch:patch_verification}
\textbf{Patch verification.}
To preserve a stable post‑patch execution state $s'$, StackPatch prohibits unbounded loops, C function calls, and dangerous instructions in patch code; any patch that violates these constraints is rejected as an unsafe patch. 
An unbounded loop would continuously occupy MCU resources, increasing system load and potentially leading to crashes. For instance, in the FreeRTOS operating system, if an unbounded loop exists within the patch code, the MCU would remain in that loop, preventing other tasks from gaining MCU access. This would compromise the system's real-time performance, causing critical tasks to miss their deadlines and resulting in abnormal system behavior. 

Certain memory locations cannot safely accommodate the patch code due to address range limitations. If a C function call is included in the patch and the function's address exceeds the calling instruction's maximum jump range, an addressing error may occur. 
For instance, in the ARMv7-M architecture, jump instructions like \texttt{bl} or \texttt{b} have a range of -16,777,216 to +16,777,214. If the vulnerable code is in Flash and the patch code is in SRAM, an invalid jump due to incorrect addressing could result in a call failure or system crash. For the purpose of stability, StackPatch avoids C function calls within the patch code. 

When adding a trampoline instruction at the patch's end, it is essential to prevent control hazards. Instructions are fetched and executed sequentially in the pipeline, so altering the control flow can cause pipeline stalls. 
For instance, if a \texttt{ldr pc, =0x08000000} instruction is placed at the end of the patch, it may already have been prefetched and decoded before execution. Modifying the \texttt{PC} register at this point can create inconsistencies between prefetched instructions and actual execution, potentially causing incorrect jumps. StackPatch addresses this by inserting a \texttt{nop} instruction before the trampoline, giving the pipeline enough time to complete instruction fetch and execution. 

In addition, operations that modify the stack state, such as \texttt{PUSH} and \texttt{POP} instructions, can complicate patching by altering the stack structure. 
In patch development, directly modifying stack states can introduce risks, so StackPatch excludes stack-altering operations to preserve the stack structure, ensuring stable and error-free patch execution. 

We measure and calculate StackPatch's delay, $T_{stackpatch}$, using the formula: 
  
\centerline{$T_{stackpatch}$ = $T_{exception}$ + $T_{dispatch}$ + $T_{patch}$,}
\noindent where $T_{exception}$ represents the fixed time of StackPatch's exception handler, $T_{dispatch}$ is the time for dispatching the patch (depending on the number of patches deployed simultaneously), and $T_{patch}$ is the patch execution time, which varies with patch code complexity. 

We detect unsafe patches by enforcing a time-threshold constraint for each real-time embedded system. Let \textit{W} be the watchdog timeout and \textit{C} the worst-case execution time of a critical task. For PLCs, we set \textit{C} to the scan time, which is one full input-execute-output cycle. We set threshold \textit{T} = \textit{W} - \textit{C}. If the measured execution time attributable to the patch causes the system to exceed \textit{T}, the patch is flagged as unsafe and rejected. If $T_{stackpatch}$ falls within the time limit, the patch is transmitted to the embedded system's update service. 
A patch is abandoned in two situations: (i) the patch is bounded but its execution time $T_{stackpatch}$ > \textit{T}; or (ii) the patch itself contains an unbounded loop, invokes C functions, or uses dangerous instructions, thereby causing $T_{stackpatch}$ to exceed \textit{W}. Any patch that could exceed the watchdog timeout and reset the MCU is rejected. 
Previous work asserts that patch programs must not contain unbounded loops and C function calls, yet provides no method for detecting such violations. In contrast, our approach introduces a concrete mechanism for identifying them.

\textbf{Update service.}
The update service receives patch information, installs patches, and sets update points. 
To receive patch information, the update service establishes a connection with the patching host through wireless protocols like Bluetooth and WiFi, or debugging channels like JTAG and UART. 
The patch information consists of a patch list, where each node includes the update point address, the memory location of the patch, the patch code, and the patch size. 
Upon receiving the patch list, the update service installs the patches in the available RAM of the target device, with each patch being accessible via its patch address. Since RAM is volatile, the patch data will be lost if the device loses power. 
The update point address can activate either software or hardware breakpoints. The update service sets the appropriate breakpoint by loading the update point address into comparison registers for hardware breakpoints, or by replacing the entry address with a trampoline instruction for software breakpoints. 

To ensure minimal disruption to the system, the update service executes patching tasks at optimal times. Control tasks (e.g., heart rate sensor tasks) are assigned the highest priority (priority 3), system stability tasks (such as delay-checking tasks) are set to priority 2, and patching tasks are given the lowest priority (priority 1). This prioritization ensures that critical tasks are not interrupted by the patching process. 

In RTOSes, StackPatch leverages the built-in idle task~\cite{freertosIdle} to perform patching. In bare-metal systems, patching can be scheduled at the end of cyclic control tasks. For example, RLPatch schedules patching between watchdog resets and scan cycles, enabling dynamic patching to address legacy vulnerabilities in PLC systems. 

\begin{algorithm}[t]
    \caption{Exception handler modification for heterogeneous MCU architectures}
    \label{alg:Exception}
    \renewcommand{\algorithmicrequire}{\textbf{Input:}}
    \renewcommand{\algorithmicensure}{\textbf{Output:}}
    \begin{algorithmic}[1]
        \REQUIRE $arch$. ($ARM, RISC-V$, $Xtensa$, etc.)  
        \ENSURE modified exception handlers    
        \IF {$arch.stack$ == $dual$-$stack$}
            \STATE $sp:$ = GetStackPointer ();
        \ENDIF
        \STATE AllocateStackSpace ($sp$);
        \STATE Save ($ra$);
        \STATE Save ($gr$);
        \IF {$arch.stack$ == $dual$-$stack$}
            \STATE UpdateStackPointer ();
        \ENDIF
        \STATE UpdatePS ();
        \STATE PatchDispatching ($sp$); 
        \IF {$arch.stack$ == $dual$-$stack$}
            \STATE $sp:$ = GetStackPointer ();
        \ENDIF
        \STATE Restore ($gr$);
        \STATE Restore ($ra$)
        \STATE FreeStackSpace ($sp$);
        \IF {$arch.stack$ == $dual$-$stack$}
            \STATE UpdateStackPointer ();
        \ENDIF
        \RETURN 
    \end{algorithmic}
\end{algorithm}

\subsection{Exception Handler Modification}
\label{stackPatch:control_flow_redirection}
Modifying the exception handler involves three main tasks: establishing the stack frame, transferring the program, and recovering the stack frame. The Algorithm~\ref{alg:Exception} describes the StackPatch modification process for exception handlers in different MCU architectures. 

\textbf{Establishing the stack frame.} 
In a dual-stack system used in target MCU architectures (e.g., ARM), the exception handler determines which stack structure is being used by selecting the appropriate stack pointer ($sp$) (lines 1-2). 
This decision is typically based on reading a register, such as \texttt{EXC\_RETURN} in the Cortex M3/M4 architecture. 
In embedded systems, dedicated registers are commonly employed to store the stack pointers. For example, in ARM architecture, the main stack pointer (\texttt{MSP}) and the process stack pointer (\texttt{PSP}) are used to store the respective stack pointers. 
The exception handler retrieves the stack pointer ($sp$) from these dedicated registers and allocates stack space (line 3) to prevent stack overflow. 

To modify the stack frame during the exception handling process, the exception handler saves the return address ($ra$)~\footnote{Typically stored in registers such as \texttt{mepc} in RISC-V or \texttt{epc} in Xtensa.} and all general-purpose registers ($gr$) to the stack (lines 4-5). 
Some MCU architectures (e.g., ARM) adopt a lazy stacking mechanism, in which exception entry performs hardware stacking that automatically saves the return address ($ra$) and a subset of general‑purpose registers ($gr$). Thus, the exception handler only needs to save the remaining registers. 
As the number and ordering of registers automatically saved by hardware vary across MCU architectures, StackPatch records the details during the construction of the $R_2$ mapping to ensure accurate stack-frame reconstruction.
Even if future silicon changes the exception entry stacking and shifts the computed stack-pointers offsets, the mapping will remain valid. 
Once all registers are saved, the dual-stack system promptly updates the stack pointer (lines 6-7) to ensure proper context saving and restoration. At this point, the stack frame $f(gr, ra)$ is fully established.

\textbf{Program switching.} 
Next, the exception handler updates the processor status registers ($PS$) (line 8), setting the processor's state and interrupt priority. 
Our exception handler's priority settings are independent of the toolchain and MCU vendor. The exception handler itself operates at the first exception level, while the patch runs at the second exception level. The status register updating process \texttt{UpdatePS} ensures that the patch inherits the same priority as the task containing the vulnerable code. StackPatch allows nested exceptions and permits higher-priority real-time tasks to preempt patch execution (the patch's priority is lower than that of real-time tasks). 
With all preparations complete, the stack pointer ($sp$) is used as a parameter to jump to the patch dispatcher (line 9). 
The patch dispatcher transfers control to the patch code, which, after execution, returns control to the exception handler, enabling the processor to restore the privileged mode subsequently.

\textbf{Recovering the stack frame.} 
Stack frame recovery mirrors the stack frame establishment process. 
In the dual-stack system, the first step is to select the stack pointer ($sp$) again (lines 10-11). 
Next, the exception handler restores the saved general-purpose registers ($gr$) and return address ($ra$) from the stack frame $f'$ (lines 12-13). 
Afterward, the allocated stack space is reclaimed using the stack pointer ($sp$) (line 14). The dual-stack system then updates the stack pointer (lines 15-16). 
Finally, after the return instruction in the exception handler (line 17), the program resumes execution with benign code $\Pi'$ in the post-patch state $s'$. 

Whether registers are saved automatically by hardware or manually by the exception handler, $ra$ is always pushed to the stack first. Restoring the program counter (PC) is deferred until all other $gr$ have been recovered, ensuring that control flow transfer occurs only after the full execution context has been safely restored.

\section{Evaluation}
\label{section:evaluation}
In this section, we conducted experiments using four embedded development boards across three MCU architectures. 
We evaluated the effectiveness of StackPatch on three embedded applications and measured its performance in terms of computational and storage resource consumption. Finally, we compare StackPatch with other state-of-the-art dynamic patching systems. 

\begin{table*}[t]
 \centering
 \small
 \begin{tabular}{l|l|l|c:c|c:c|c:c}
\hline
\multirow{3}{*}{ \textbf{CVE-ID} } &\multirow{3}{*}{\textbf{OS / Lib}} &\multicolumn{1}{l|}{\multirow{3}{*}{\textbf{Vulnerability type}}}&\multicolumn{2}{c|}{\textbf{STM32F401RE (84MHz)}}  & \multicolumn{2}{c|}{\textbf{GD32VF103 (108MHz)}}  & \multicolumn{2}{c}{\textbf{ESP32S3 (240MHz)}}  \\ \cline{4-9} 
  &&& \makecell[c]{ ExecTime \\ (MCU cycles)} & \makecell[c]{ MemUsage \\ (bytes)}& \makecell[c]{ ExecTime\\ (MCU cycles)} &\makecell[c]{ MemUsage \\  (bytes)}  & \makecell[c]{ExecTime \\(MCU cycles)}&\makecell[c]{ MemUsage \\  (bytes)}  \\  \hline

CVE-2020-10021 &Zephyr OS&Out-of-Bounds Write & 20 & 42 & 18& 36  & 26& 40 \\
CVE-2020-10023 &Zephyr OS&Buffer Overflow         & 12& 24 & 14& 16 & 17& 24 \\
CVE-2020-10024 &Zephyr OS&Instruction Misuse  & 9  & 28 & 12& 62 & 35& 44 \\
CVE-2020-10028 &Zephyr OS&Lack Sanity Checking& 15& 36 & 19& 28 & 42& 82 \\
CVE-2020-10062 &Zephyr OS&Logical Bug         & 13  & 42 & 18& 32 & 22& 28 \\
CVE-2020-10063 &Zephyr OS&Integer Overflow    & 19  & 44 & 22& 42 & 26& 52\\\hline

CVE-2018-16524 &FreeRTOS&Lack Validation      & 15 & 32 & 14& 28 & 21& 32 \\
CVE-2018-16528 &FreeRTOS&State Confusion      & 17 + 19 & 32 + 40 & 12 + 14& 22 + 28& 41 + 43& 32 + 38 \\
CVE-2018-16603 &FreeRTOS&Out-of-Bounds Read   & 22& 52 & 31& 42 & 38& 50 \\ \hline

CVE-2017-2784 &mbedTLS&Invalid Free           & 6 & 12 & 2& 8   & 24& 28 \\
CVE-2020-17443 &AMNESIA33&Lack Packet Checks  & 16 & 40 & 16& 30 & 33& 67 \\
CVE-2020-17445 &AMNESIA33&Lack Option Checks  & 18& 50& 18& 34 & 24& 40\\ \hline
Average&&&15& 39	& 16	& 34& 30 & 46
 \\ \hline
  \end{tabular}
   \caption{The performance of StackPatch on STM32F401RE (ARM Cortex-M4), GD32VF103 (RISC-V32), and ESP32S3 (Xtensa LX7) boards across various vulnerability types. Notable, CVE-2018-16528 required two hot patches for resolution.}
   \label{tab:generality}
\end{table*}

\subsection{Experimental Setup}
\label{evaluation:environmental_setup}
In this study, the StackPatch was evaluated across four embedded MCU boards with diverse architectures and capabilities: the nRF52840 (ARM Cortex-M4), STM32F401RE (ARM Cortex-M4), GD32VF103 (RISC-V32), and ESP32S3 (Xtensa LX7). 
These boards were selected to represent three common embedded scenarios: industrial automation, smart medical devices, and network services. 
For these scenarios, we evaluate a Soft PLC, a heart-rate monitor, and an HTTP/CoAP handler as the respective applications. The three applications span distinct workload classes: real-time control, real-time perception, and network throughput. 
The Soft PLC and the heart-rate monitor have hard real-time deadlines, where missing a deadline can cause system failure; whereas the HTTP/CoAP handler is best effort and constrained mainly by throughput. 
Each of the four development boards was configured with the necessary peripherals to support the target applications. 
The nRF52840 board, equipped with an ARM Cortex-M4 processor, was utilized in the heart rate monitoring experiment, leveraging its low-power capabilities and Bluetooth 5.0 support. 
This board was connected to a photoplethysmography (PPG) sensor to detect blood flow through changes in light absorption via red and infrared LEDs~\cite{Allen}, enabling real-time heart rate detection and data transmission to a host computer. 
Similarly, the GD32VF103 board, based on the open-source RISC-V32 architecture, was also used in the heart rate monitoring setup, where it interfaced with the same PPG sensor. This board offered a cost-effective solution for the experiment, targeting resource-constrained environments typical in medical devices.

For the Soft PLC application, the STM32F401RE board was selected due to its high-performance processing capabilities and flexible peripheral integration, which are essential for industrial control systems. 
The setup consisted of an X-Nucleo-PLC01A1 industrial I/O expansion board, responsible for handling physical input and output signals, and an X-Nucleo-IDW01M1 wireless board for receiving patch data. 
The STM32F401RE board ran bare-metal firmware built from STMicroelectronics' official \texttt{FP\_IND\_PLCWIFI1} package~\cite{STM32CubeFunctionPack}. 
The Soft PLC system operates on a \texttt{systick} interrupt to perform a cyclic scan every 0.012 ms, continuously monitoring inputs and controlling outputs such as LED indicators. 
A watchdog timer was used to ensure system reliability, automatically resetting the system in the event of interference or errors. 

We evaluated StackPatch on an ESP32S3 board (Xtensa LX7 processor) by patching network servers. 
Two applications were developed: a Constrained Application Protocol (CoAP) server and a Hypertext Transfer Protocol (HTTP) server, both built on the ESP-IDF framework~\cite{espidf}, which integrates FreeRTOS. 
These servers were used to evaluate network throughput in IoT scenarios, where StackPatch applied security updates to the system without interrupting ongoing operations. 

Vulnerabilities were collected from widely-used RTOSes such as Amazon FreeRTOS and Zephyr OS, along with several libraries including PicoTCP~\cite{picotcp}, WolfSSL~\cite{wolfSSL}, MbedTLS~\cite{MbedTLS}, and AMNESIA33~\cite{AMNESIA33}. 
A total of 107 vulnerabilities (CVEs) were identified, with 63 classified as HIGH (CVSS 7.0–8.9) and 25 as CRITICAL (CVSS 9.0–10.0). 
These vulnerabilities were selected based on three criteria: a CVSS score above 5.0, the availability of official patch source code to facilitate StackPatch generation, and the inclusion of vulnerabilities that could also be addressed by alternative patching methods for performance comparison. 
The identified vulnerabilities were used to test StackPatch's ability to generate patches and secure the embedded systems in real-time.

\begin{figure*}[t]
    \centering
    \includegraphics[width=1\textwidth]{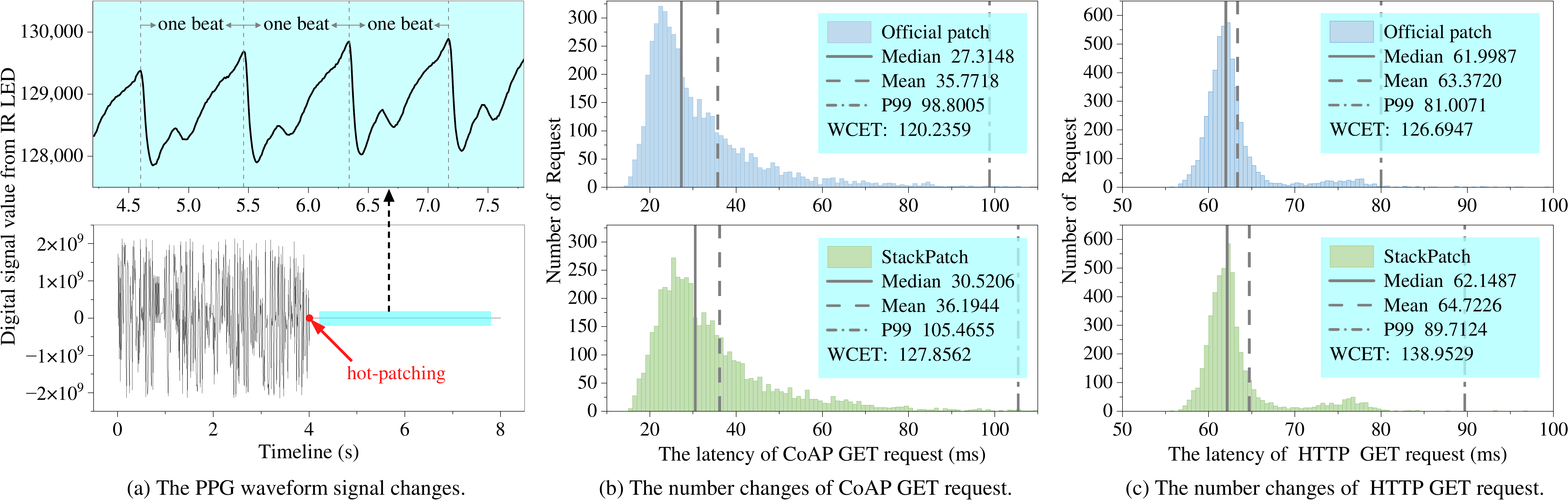}
    \caption{Changes in measurements for hot-patching heart rate monitor and network service applications.}
    \Description{Changes in measurements for hot-patching heart rate monitor and network service applications.}
    \label{fig:availability}
\end{figure*}

\subsection{Effectiveness of StackPatch}
\textbf{Vulnerability patching.} 
StackPatch generated secure patches for 107 vulnerabilities, which were applied to four MCU boards: STM32F401RE, nRF52840, GD32VF103, and ESP32S3. 
Using a modified exception handler, 102 vulnerabilities in RTOSes and libraries (Table~\ref{tab:CVEtable} in Appendix~\ref{appendix:repair_effectiveness}) were successfully patched without system downtime or interference with applications. 
To evaluate the impact of compiler optimization on StackPatch, we selected 12 CVEs (Table~\ref{tab:generality}) that span the most common categories among the 107 embedded CVEs, using one exemplar per category. 
For each CVE, we built targets at four optimization levels: -O0, -O1, -O2, and -O3. Across all levels, StackPatch consistently located identical update points and generated semantically equivalent patches. 
All vulnerabilities were successfully repaired at every optimization level (12/12, 100\%). These results indicate that compiler optimization has no impact on update-point localization or patch generation.

It is worth noting that the official patch for CVE-2019-18840 introduced five boundary checks, which StackPatch successfully fixed. However, RapidPatch and AutoPatch were unable to resolve the issue. RapidPatch failed because the boundary checks are spread across a large function, and its filter patch can only check the boundary once, at the function's entry point. 
To fix the vulnerability, the patch would need to cover the entire function, but the function's size prevented RapidPatch from creating a successful patch. 
AutoPatch also failed for two reasons: first, the large offset between its instrumentation point and the final update point resulted in five large patches, and second, its weakest precondition reasoning algorithm cannot correctly construct the patch~\cite{Zhengzi}. 
StackPatch could not repair five vulnerabilities: CVE-2020-10064, CVE-2020-13598, CVE-2022-2741, CVE-2020-10072, and CVE-2024-11263. 
In each case, the vendor patch contains dozens of code edits and modifies four to seven macro definitions. 
The last two vulnerabilities also require modifications to the \texttt{CMakeLists} file, along with numerous code modifications. 
Fixing these vulnerabilities with StackPatch would require tens or even hundreds of update points, placing significant strain on the system's limited resources. 

\textbf{Soft PLC patching.} 
To repair a buffer overflow vulnerability CVE-2020-10023 in the soft PLC system, StackPatch's update service was scheduled during idle cycles. 
After deployment, the system ran continuously without a restart, indicating that StackPatch is applicable to bare-metal systems. 

\begin{table}[t]
    \centering 
    \small
 \setlength{\tabcolsep}{1.5mm}{
    \begin{tabular}{l|c|c|c|c}
    \hline  
          \multirow{2}{*}{ \textbf{MCU boards}}&\multirow{2}{*}{\textbf{Frequency}}&\multicolumn{3}{c}{\textbf{Mean execution time (MCU cycles)}} \\\cline{3-5}
             &   &$T_{exception}$&$T_{dispatch}$&$T_{stackpatch}$ \\ \hline 
         
         nRF52840&64MHz    & 61 &31 $\sim$ 198 & 92 $\sim$ 259 \\ \hline 
         STM32F401RE&84MHz & 61 &31 $\sim$ 198 & 92 $\sim$ 259 \\ \hline
         GD32VF103&108MHz  & 90 &38 $\sim$ 141 & 128 $\sim$ 231 \\ \hline
         ESP32S3&240MHz    & 69 &24 $\sim$ 191 & 93 $\sim$ 260  \\\hline
    \end{tabular}}
    \caption{The mean execution time of StackPatch across four development boards.}
    \label{tab:table4}
\end{table}

\textbf{Heart rate monitor patching.} 
In the heart rate sampling application, the out-of-bounds read vulnerability CVE-2018-16601 caused memory errors and disrupted PPG waveforms. StackPatch dynamically applied a patch after update points were installed, restoring correct PPG signals (Figure~\ref{fig:availability}.a). 
The system remained stable, with LED blinking and accurate heart rate monitoring confirmed by the magnified view at the top of Figure~\ref{fig:availability} (a). 
StackPatch ensured security and stability without interrupting the application.

\textbf{Network service patching.}
To assess the worst-case latency and runtime overhead, we evaluated StackPatch using its hook-based trigger mechanism. 
We compared an officially patched build with a StackPatch‑patched firmware. Using FreeRTOS on an ESP32‑S3 (240 MHz) board, we reproduced four vulnerabilities (CVE‑2018‑19523, CVE‑2018‑16524, CVE‑2018‑16528, CVE‑2018‑16601) and applied the fixes either at the official patch or via StackPatch. 
For the StackPatch configuration, we instrumented 1,535 functions (34$\%$ of all functions) across FreeRTOS and its sub-libraries. 
We then ran a high‑load network service and issued 5,000 consecutive HTTP GET and CoAP GET requests from the host, sustaining about 28 HTTP and 15 CoAP requests per second. 
Under this workload, StackPatch successfully patched all four vulnerabilities in parallel. 
For CoAP, StackPatch increased average latency by 1.2$\%$ ( (36.1944-35.7718)/ 35.7718), the 99th percentile (P99) latency by 6.7$\%$, and the worst-case execution time (WCET) by 6.3\% (Figure~\ref{fig:availability}.b). 
For HTTP, average latency rose by 2.1$\%$, the P99 latency by 10.7$\%$, and the WCET by 9.7\% (Figure~\ref{fig:availability}.c). 
These results indicate StackPatch's low overhead and feasibility for embedded systems. 

\begin{table*}[t]
 \centering
 \small
\begin{tabular}{l|l|l|l|l|l|l|l|l}\hline
\makecell[l]{ \textbf{System}}  &\textbf{Target} & \makecell[c]{ \textbf{Patch} \\ \textbf{gran.} } & \textbf{Update strategy} &\makecell[c]{ \textbf{Multi}-\textbf{arch}\\ \textbf{support}} &\textbf{Patch language} &\makecell[c]{\textbf{Data-structure} \\ \textbf{change support}} & \makecell[c]{ \textbf{Runtime} \\ \textbf{MemUsage} } &\makecell[c]{ \textbf{Mean} \\ \textbf{patch} \textbf{size}} \\\hline
UPStare &Apps &  Function & Signal mech. & Support & C source code & Value change & 64 KB & - \\\hline
kpatch &Kernel   &  Function & System calls &Support& C source code & Value change & 20 MB  & - \\\hline
HERA & \multirow{5}{*}{\makecell[l]{ Embedded \\ Firmware} } & Basic block & FPB unit &Unsupport& Assembly code  & Unsupport & - & - \\
RapidPatch& & Basic block & eBPF VM             &Support& eBPF code     & Unsupport & 18 KB& 102 bytes   \\
AutoPatch&  & Basic block & Instrumentation     &Support& LLVM IR       & Unsupport & - & 535 bytes   \\\
RLPatch  &  & Basic block & Exception mech. &Unsupport& Assembly code & Unsupport & 2 KB & 872 bytes   \\
StackPatch& & Basic block & Exception mech. &Support& C source code & Value and size change & 0.7 KB & 40 bytes    \\\hline
\end{tabular}
   \caption{Comparison of StackPatch with existing dynamic patching systems.}
   \label{tab:comparison}
\end{table*}

\subsection{Performance Evaluation}
\label{evaluation:performance_evaulation}
\textbf{Computational resource consumption.} 
Computational overhead is attributed to three components: the exception handler, the patch dispatcher, and the hot patch. 
The exception handler consists of a fixed number of processor instructions, resulting in constant execution time. 
For statistical reliability, each measurement was repeated 300 times and the mean was reported. 
Using a high-precision counter synchronized with the MCU frequency, we measured exception handler times ($T_{exception}$) for the four devices as 61, 61, 90, and 69 MCU cycles, respectively (Table~\ref{tab:table4}). 

Since the components of StackPatch operate consecutively, it is not possible to directly measure the patch dispatcher time ($T_{dispatch}$). 
Instead, we calculated it by subtracting $T_{exception}$ from the total execution time ($T_{stackpatch}$), measured using software breakpoints. 
Specifically, we inserted a software breakpoint instruction at the vulnerability entry point and recorded the counter values before and after its execution to calculate the number of MCU cycles, thereby determining $T_{stackpatch}$. 
The last two columns of Table~\ref{tab:table4} display $T_{dispatch}$ and $T_{stackpatch}$ for scheduling 1$\sim$64 patches in the patch dispatcher and executing an empty patch. Note that the patch dispatcher can handle up to 64 patches, limited by the memory resources of embedded devices. 
On ARM Cortex-M4 (nRF52840 and STM32F401RE boards), the exception time ($T_{exception}$) in MCU cycles is the lowest because lazy stacking saves part of the register operations in hardware, reducing instructions in the exception handler. 
On RISC-V (GD32VF103 board), $T_{exception}$ is the highest because there is no hardware stacking and the handler must save all general-purpose registers in software, which lengthens the handler. 
Across the 64 patches, the worst-case $T_{stackpatch}$ is governed primarily by the architecture's exception-entry behavior and instruction execution efficiency. ARM and Xtensa (ESP32S3 board) yield the lowest instruction efficiency, whereas RISC-V yields the highest instruction efficiency. 

To evaluate hot patch execution time, we selected 12 CVEs from a corpus of 107 embedded vulnerabilities (Table~\ref{tab:generality}). 
Each chosen CVE represents a prevalent category in the corpus, ensuring that the benchmark covers the dominant vulnerability types seen in practice. 
On the STM32F401RE board, we tested three triggering mechanisms (hardware breakpoints, software breakpoints, and hooks), achieving an average patch latency of 0.18 $\mu$s (15 MCU cycles). 
On the GD32VF103 and ESP32S3 boards, we used software breakpoints and hooks, achieving average patch latencies of 0.15 $\mu$s (16 MCU cycles) and 0.13 $\mu$s (30 MCU cycles), respectively. 
Although the GD32VF103 supports up to four hardware breakpoints in Flash~\cite{gd32}, the official documentation lacks implementation details, so hardware breakpoints were not utilized on this device. 
Although higher-frequency devices generally exhibit lower patch execution times, architectural differences also play a significant role.

\textbf{Storage resource consumption.}
StackPatch's memory overhead is minimal. Across ARM, RISC-V, and Xtensa architectures, modifying the exception handler required only 82 bytes, 160 bytes, and 51 bytes, respectively. 
Single patch code sizes of 12 selected CVEs averaged 39 bytes, 34 bytes, and 46 bytes for STM32F401RE, GD32VF103, and ESP32S3 boards, respectively (Table~\ref{tab:generality}). 
On the nRF52840 board, testing 102 vulnerabilities yields a mean patch payload of 40 bytes, and memory usage was consistent across applications on this architecture. 

The update service comprises only a few dozen lines of C code. The StackPatch runtime incurs a footprint of 0.5 KB in flash memory and 0.2 KB in SRAM. 

\subsection{Comparison}
Table~\ref{tab:comparison} compares StackPatch with UpStare, kpatch, HERA, RapidPatch, AutoPatch, and RLPatch along four dimensions: multi-architecture support, diversity of update strategy, deployment footprint, and capability to handle high-complexity patches. 

\textbf{Multi-architecture support.}
UpStare is largely architecture-agnostic because it applies source-to-source transformations in CIL~\cite{necula2002cil}, a high-level C intermediate representation. Kpatch attains broad hardware coverage by relying on Linux kernel facilities such as \texttt{ftrace}, which are implemented across many architectures. 
HERA relies on the FPB unit, limiting it to ARM Cortex-M3/M4 architectures. RLPatch is specific to ARM-based PLCs~\cite{MingZhou}, while RapidPatch and AutoPatch leverage the eBPF virtual machine and LLVM compiler, respectively, to support multiple architectures. 
StackPatch, by reconstructing the stack frame and rewriting the exception handler, supports heterogeneous MCU architectures. 

\textbf{Diversity of update strategy.}
In full operating system environments, UpStare and kpatch can rely on the standard kernel signal mechanism to trigger updates, so multiple trigger types are unnecessary. 
Embedded firmware, by contrast, must cope with heterogeneous hardware and tight resource budgets, and thus benefits from a broader set of triggers. 
Support across embedded approaches is uneven: HERA relies solely on FPB-based address remapping; AutoPatch depends on extensive instrumentation; RLPatch supports both hardware and software breakpoints; RapidPatch combines hardware breakpoints with instrumentation. 
StackPatch provides three trigger mechanisms: hardware breakpoints, software breakpoints, and compile-time hooks. 
It selects the mechanism according to memory type and patch density: with only a few vulnerabilities, it prefers hardware breakpoints; when many sites must be patched, it instruments code in flash and uses software breakpoints for code in SRAM, conserving limited hardware breakpoints while keeping latency low. 

\begin{figure}[t]
    \centering
    \includegraphics[width=0.475\textwidth]{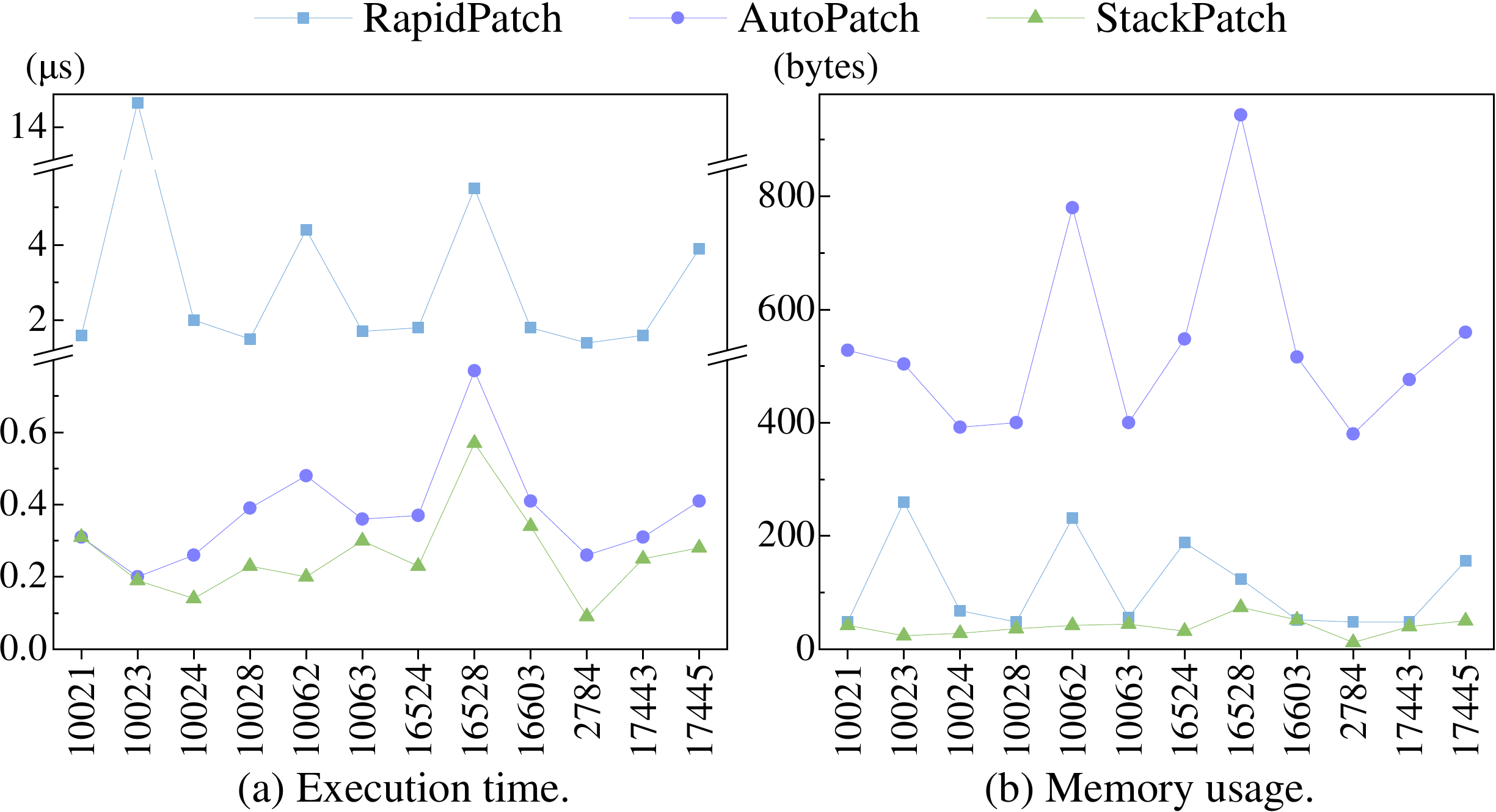}
    \caption{Comparison of execution time (a) and memory usage (b) for 12 CVEs (truncate the year) of varying vulnerability types across StackPatch, AutoPatch, and RapidPatch on the nRF52840 board (64 MHz).}
    \Description{Comparison.}
    \label{fig:nrfcompare}
\end{figure}

\textbf{Deployment footprint.}
Among systems that report runtime memory usage, StackPatch has the smallest footprint (0.7 KB). 
In contrast, UpStare and kpatch target general-purpose operating systems and require substantial runtime support, which leads to much higher memory consumption. 
Given the heterogeneity of MCUs and tight resource budgets in embedded firmware, we focus the comparison on patching systems designed for embedded platforms. 
Because HERA and RLPatch lack public implementations, we limit the direct comparison to AutoPatch and RapidPatch. We evaluate the three systems on the nRF52840 board using 12 CVEs that span multiple vulnerability categories. 
As shown in Figure~\ref{fig:nrfcompare}, StackPatch consistently yields the lowest execution latency and memory usage per patch, owing to its fine-grained update point selection at the instruction level. 
RapidPatch has the highest execution time because it executes patches inside an eBPF virtual machine, whereas AutoPatch has the largest memory footprint due to its file-based patch format and pre-positioned instrumentation at the basic-block level. 

We also scheduled 1$\sim$5 patches in the patch dispatcher to measure the total time required to fix CVE-2018-16601. RapidPatch and AutoPatch required 4.1$\sim$5.38 $\mu$s, and  3.0$\sim$4.37 $\mu$s, respectively. StackPatch required 2.01$\sim$2.06 $\mu$s to repair this vulnerability using hardware breakpoints, 2.52$\sim$2.83 $\mu$s using software breakpoints, and 2.11$\sim$2.42 $\mu$s using instrumentation. 
The differences among StackPatch's mechanisms are negligible, and its average total patching time is over 50\% faster than AutoPatch and more than twice as fast as RapidPatch.
The additional overhead in RapidPatch stems from the virtual machine, while AutoPatch incurs overhead due to extensive instrumentation. HERA and RLPatch, using assembly-based patches, likely have minimal overhead but require manually written assembly code.

\textbf{Capability to handle high-complexity patches.}
Both HERA and RLPatch require manually written assembly patches, imposing substantial demands on RTOS developers who must work across multiple instruction sets. 
In addition, creating hot patches in assembly increases the risk of errors, complicating the development process. 
In contrast, RapidPatch uses third-party eBPF bytecode, requiring developers to write patches in C and then convert them to eBPF code. This extra step is time-consuming, particularly for large-scale CVE fixes.

AutoPatch uses weakest-precondition reasoning to synthesize hot patches but lacks built-in correctness verification, requiring manual review of patches in LLVM IR [49].
Its reliance on numerous instrumentation points also imposes substantial MCU and memory overhead, limiting practicality on resource‑constrained embedded systems. 
Furthermore, these embedded-oriented systems cannot synthesize complex patches that modify global variables or change macro definitions. 
Although UpStare and kpatch introduce type transformers and shadow variables to accommodate some structure updates, they still cannot safely apply patches that change the size of a data structure. 

StackPatch addresses these challenges by mapping variables to stack pointers, simplifying patch implementation. RTOS developers can write patches directly in C without the need to convert them into third-party languages, lowering the technical barrier and enhancing the accuracy and safety of the patching process. 
Across 102 CVEs, StackPatch completed vulnerability localization, update‑point selection, patch generation, and offline verification within 10 minutes in more than 90\% of cases. The slowest case was CVE‑2020‑10022, which required changes to global and local variables as well as macro definitions; the end‑to‑end process finished in 20 minutes. 
StackPatch can effectively handle 12 common RTOS vulnerability types (Table~\ref{tab:generality}), including out-of-bounds writes, out-of-bounds reads, buffer overflows, and logical errors. It also supports complex scenarios, such as data structure modifications.

\section{Related Work}
\label{section:related_work}
Hot patching, also called dynamic vulnerability patching, enables software to be modified or updated while it is running, which fulfills high availability requirements. 

\textbf{Hot patching on traditional systems.}
Prior work, such as POLUS~\cite{Haibo}, UpStare~\cite{Makris}, Javelus~\cite{Tianxiao}, LIBBANDAID~\cite{YueAG}, and OSSPATCHER~\cite{Duan}, focuses on hot patching for traditional user‑space applications. 
UpStare is the closest in spirit: it reconstructs stack frames using signal handlers and runtime dynamic instrumentation to modify functions that are active on the stack. This model does not transfer to embedded systems, because most RTOSes and bare‑metal firmware provide neither signal handling nor user‑mode context switching, and thus signal handlers cannot be delivered. 
Moreover, kernel patches must be applied in place at the faulting instruction to preserve system state, whereas UpStare operates at function entries and exits and cannot handle unexpected out‑of‑bounds accesses or illegal instructions. 
StackPatch addresses these constraints by combining hardware exceptions with instruction‑level reconstruction, enabling precise localization and repair of compiler‑optimized instructions along the exception path in embedded kernels while keeping runtime overhead low. 

Prior work has introduced kernel hot‑patching for platforms such as Linux~\cite{Arnold, Kpatch, kgraft, MakrisD, Zhenyu} and Android~\cite{YueAA, Zhengzi, Xuewen}, but these approaches incur substantial resource overhead and require downtime, making them unsuitable for real‑time embedded systems that demand bounded latency and continuous operation. 

\textbf{Hot patching on real-time embedded systems.}
Recent years have seen progress in hot patching for real-time embedded systems.
Wahler \etal~\cite{WahlerDS,wahlerNL,wahlerDF} proposed a state transfer model where components communicate through channel-based message-passing mechanisms and rely on memory protection to ensure separation. State transfer represents the shared memory space of an updated component, but its main limitation is that the transfer must fit within one cycle.
Simon \etal~\cite{holmbacka} introduced a mechanism to link and relink FreeRTOS tasks at runtime dynamically, but it can only update user-level tasks, not the RTOS kernel~\cite{YiHe}.
HERA~\cite{Niesler} utilized the FPB unit to atomically redirect control flow to hot patches. Due to its dependency on processors' deprecated features, HERA is only applicable on ARM Cortex M3/M4-based devices with a limited number of patches.
To address the issue of dynamic vulnerability patching in control programs, ICSPatch~\cite{Rajput} reverse-engineers the binary structure of control programs and uses data dependency graphs (DDG) to locate and fix vulnerabilities within the control programs themselves, rather than at the OS or firmware level.
To handle dynamic vulnerability patching of closed-source industrial controller firmware, RLPatch~\cite{MingZhou} leverages the processor's built-in exception handling mechanism to fix firmware vulnerabilities dynamically and selects idle times during the PLC's scan cycle to switch programs. 

\textbf{Hot patching for heterogeneous embedded devices.}
RapidPatch~\cite{YiHe} introduces the eBPF virtual machine to generate universal patches for heterogeneous embedded devices. However, eBPF VM incurs memory overhead and only provides limited tail calls for extending instructions. Our proposed StackPatch does not have this limitation. In the framework of StackPatch, we show that stack reconstruction with built-in exception-handling mechanisms of real-time embedded systems can achieve high flexibility and versatility in patching heterogeneous embedded systems with minimal resource consumption and overhead.

\section{Limitations}
\label{section:limitation}
\textbf{Complex macros and configuration changes.}
StackPatch does not robustly support repairing multiple multi-line macros or complex expressions in a single update (for example, \texttt{a = (b > c)? $func\_1$(): $func\_2$()}). 
Each macro expansion site requires its own update point; when a macro expands at many sites, the number of concurrent update points can grow to dozens, causing code-space and bookkeeping pressure. 
Similarly, changes to build or configuration files (such as \texttt{CMakeLists}) typically coincide with widespread source modifications, which could demand a large set of coordinated update points and patches. 
To the best of our knowledge, no existing technique can handle either scenario on resource-constrained embedded systems. 

\textbf{No support for multi-core systems.}
StackPatch is primarily suited for embedded systems with low-power, single-core processors, such as those in PLCs and medical devices. 

\textbf{Lack of automatic pipeline.}
In StackPatch, both vulnerability localization and update point determination are done manually to ensure the accuracy and reliability of hot patches. 

\section{Discussion}
\label{section:discussion}
In comparison with prior embedded hot patching systems, StackPatch repairs complex vulnerabilities, including data-structure changes to local and global variables and updates to macro definitions. 
It also supports flexible cross-architecture update strategies and lightweight deployment, making it suitable for diverse embedded scenarios. 

StackPatch targets low‑power, single‑core embedded platforms. 
In multicore settings, per‑core caches can diverge during a hot patch, delaying or preventing the fix from taking effect. 
This risk can be mitigated by invoking cache‑coherence mechanisms (e.g., snooping or directory‑based protocols) to refresh all cores. 
Extending StackPatch with explicit coherence handling for automotive, robotics, and high‑end consumer systems is our future work.

Function‑level updaters such as UpStare and kpatch do not scale to large functions on storage‑constrained devices, because replacing an entire function often exceeds available space. 
StackPatch operates at the instruction level. It splits a function into small code fragments and treats each fragment as an independent update unit, allowing triggers to be placed anywhere within the function and enabling reliable hot patching under tight resource budgets.

Accurate vulnerability localization is the prerequisite for any automatic determination of update points. 
At present, this localization step remains manual in both offline and dynamic patching. There is no automated method at the machine‑instruction level; even at the source level, systems such as Grace~\cite{GRACE}, DEEPRL4FL~\cite{DEEPRL4FL}, TRANSFER~\cite{TRANSFER}, and LLMAO~\cite{Yang2024LLMfTFL} report only 22.3\% accuracy, with a top‑5 rate of 46.3\% (the correct buggy line appears among the five highest‑scoring candidates). 
Recent work, AutoPatch, also requires vulnerabilities to be localized manually. 

A large‑scale empirical study by Heyden \etal~\cite{Hayden2012EvaluatingDS} reveals that, for real‑time systems, automatically chosen update points frequently fail due to version mismatches that cause timeouts. 
RLPatch~\cite{MingZhou} and our results show that manual update‑point selection yields minimal‑latency updates. 
Since selection is a one‑time task, the associated manual effort is justifiable. 

\section{Conclusions}
\label{section:conclusion}
The widespread deployment of heterogeneous embedded devices in mission-critical tasks requires an architecture-agnostic and timely vulnerability patching mechanism. 
In this paper, we proposed StackPatch, the first stack-frame reconstruction-based hot patching framework for embedded devices, which generates and inserts hot patches with less overhead and zero downtime. 
Specifically, StackPatch generates well-tailored hot patches based on static analysis and stack frame remapping. The generated patch instructions are triggered by control flow redirection based on the processor's built-in exception-handling mechanisms.
We conducted comprehensive experiments to evaluate the effectiveness and performance of StackPatch on embedded devices with three major microcontroller (MCU) architectures: ARM, RISC-V, and Xtensa.
The evaluation results show that StackPatch successfully fixes 102 publicly disclosed vulnerabilities in real-time operating systems (RTOS). 
We applied patching in medical devices, soft programmable logic controllers (PLCs), and network services, and the additional overhead incurred by StackPatch is only within 260 MCU clock cycles.

\section*{Availability}

We have made the relevant programs of StackPatch and experiment data publicly available at \url{https://github.com/xumesang/StackPatch}.

\begin{acks}
We would like to thank the anonymous reviewers for their insightful and detailed feedback. 
This work was supported in part by the National Natural Science Foundation of China (NSFC) under Grant Nos. 62402225 and 92467201, and by the U.S. Office of Naval Research under Grant No. N00014-23-1-2158. 
\end{acks}

\bibliographystyle{ACM-Reference-Format}
\balance
\bibliography{ref}


\begin{thebibliography}{57}


\ifx \showCODEN    \undefined \def \showCODEN     #1{\unskip}     \fi
\ifx \showISBNx    \undefined \def \showISBNx     #1{\unskip}     \fi
\ifx \showISBNxiii \undefined \def \showISBNxiii  #1{\unskip}     \fi
\ifx \showISSN     \undefined \def \showISSN      #1{\unskip}     \fi
\ifx \showLCCN     \undefined \def \showLCCN      #1{\unskip}     \fi
\ifx \shownote     \undefined \def \shownote      #1{#1}          \fi
\ifx \showarticletitle \undefined \def \showarticletitle #1{#1}   \fi
\ifx \showURL      \undefined \def \showURL       {\relax}        \fi
\providecommand\bibfield[2]{#2}
\providecommand\bibinfo[2]{#2}
\providecommand\natexlab[1]{#1}
\providecommand\showeprint[2][]{arXiv:#2}

\bibitem[Agency(2019)]%
        {Ghidra}
\bibfield{author}{\bibinfo{person}{National~Security Agency}.} \bibinfo{year}{2019}\natexlab{}.
\newblock \bibinfo{title}{{G}hidra}.
\newblock \bibinfo{howpublished}{\url{https://ghidra-sre.org/}}.
\newblock


\bibitem[Allen(2007)]%
        {Allen}
\bibfield{author}{\bibinfo{person}{John Allen}.} \bibinfo{year}{2007}\natexlab{}.
\newblock \showarticletitle{Photoplethysmography and its application in clinical physiological measurement}.
\newblock \bibinfo{journal}{\emph{Physiological measurement}} \bibinfo{volume}{28}, \bibinfo{number}{3} (\bibinfo{year}{2007}), \bibinfo{pages}{R1}.
\newblock


\bibitem[Android(2025)]%
        {androidAB}
\bibfield{author}{\bibinfo{person}{Android}.} \bibinfo{year}{2025}\natexlab{}.
\newblock \bibinfo{title}{{A}/{B} (seamless) {S}ystem {U}pdates in {A}ndroid {D}evices.}
\newblock \bibinfo{howpublished}{\url{https://source.android.com/docs/core/ota/ab}}.
\newblock


\bibitem[ARM(2025)]%
        {fpb}
\bibfield{author}{\bibinfo{person}{ARM}.} \bibinfo{year}{2025}\natexlab{}.
\newblock \bibinfo{title}{{T}he {F}lash {P}atch and {B}reakpoint {U}nit ({F}{P}{B}) in {ARM} {C}ortex {M}3/{M}4.}
\newblock \bibinfo{howpublished}{\url{https://developer.arm.com/documentation/ddi0337/h/debug/about-the-flash-patch-and-breakpoint-unit--fpb-}}.
\newblock


\bibitem[Arnold and Kaashoek(2009)]%
        {Arnold}
\bibfield{author}{\bibinfo{person}{Jeff Arnold} {and} \bibinfo{person}{M.~Frans Kaashoek}.} \bibinfo{year}{2009}\natexlab{}.
\newblock \showarticletitle{Ksplice: automatic rebootless kernel updates}. In \bibinfo{booktitle}{\emph{Proceedings of the 4th ACM European Conference on Computer Systems}} (Nuremberg, Germany) \emph{(\bibinfo{series}{EuroSys '09})}. \bibinfo{publisher}{Association for Computing Machinery}, \bibinfo{address}{New York, NY, USA}, \bibinfo{pages}{187–198}.
\newblock
\showISBNx{9781605584829}
\href{https://doi.org/10.1145/1519065.1519085}{doi:\nolinkurl{10.1145/1519065.1519085}}


\bibitem[Belgium(2025)]%
        {picotcp}
\bibfield{author}{\bibinfo{person}{Altran~EESY Belgium}.} \bibinfo{year}{2025}\natexlab{}.
\newblock \bibinfo{title}{{T}he {P}ico{T}{C}{P} {L}ibrary.}
\newblock \bibinfo{howpublished}{\url{https://github.com/tass-belgium/picotcp}}.
\newblock


\bibitem[Chen et~al\mbox{.}(2007)]%
        {Haibo}
\bibfield{author}{\bibinfo{person}{Haibo Chen}, \bibinfo{person}{Jie Yu}, \bibinfo{person}{Rong Chen}, \bibinfo{person}{Binyu Zang}, {and} \bibinfo{person}{Pen-Chung Yew}.} \bibinfo{year}{2007}\natexlab{}.
\newblock \showarticletitle{POLUS: A POwerful Live Updating System}. In \bibinfo{booktitle}{\emph{Proceedings of the 29th International Conference on Software Engineering}} \emph{(\bibinfo{series}{ICSE '07})}. \bibinfo{publisher}{IEEE Computer Society}, \bibinfo{address}{USA}, \bibinfo{pages}{271–281}.
\newblock
\showISBNx{0769528287}
\href{https://doi.org/10.1109/ICSE.2007.65}{doi:\nolinkurl{10.1109/ICSE.2007.65}}


\bibitem[Chen et~al\mbox{.}(2017)]%
        {YueAA}
\bibfield{author}{\bibinfo{person}{Yue Chen}, \bibinfo{person}{Yulong Zhang}, \bibinfo{person}{Zhi Wang}, \bibinfo{person}{Liangzhao Xia}, \bibinfo{person}{Chenfu Bao}, {and} \bibinfo{person}{Tao Wei}.} \bibinfo{year}{2017}\natexlab{}.
\newblock \showarticletitle{Adaptive android kernel live patching}. In \bibinfo{booktitle}{\emph{Proceedings of the 26th USENIX Conference on Security Symposium}} (Vancouver, BC, Canada) \emph{(\bibinfo{series}{SEC'17})}. \bibinfo{publisher}{USENIX Association}, \bibinfo{address}{USA}, \bibinfo{pages}{1253–1270}.
\newblock
\showISBNx{9781931971409}


\bibitem[Dong et~al\mbox{.}(2013)]%
        {Dong}
\bibfield{author}{\bibinfo{person}{Wei Dong}, \bibinfo{person}{Biyuan Mo}, \bibinfo{person}{Chao Huang}, \bibinfo{person}{Yunhao Liu}, {and} \bibinfo{person}{Chun Chen}.} \bibinfo{year}{2013}\natexlab{}.
\newblock \showarticletitle{{R3:} Optimizing relocatable code for efficient reprogramming in networked embedded systems}. In \bibinfo{booktitle}{\emph{Proceedings of the {IEEE} {INFOCOM} 2013, Turin, Italy, April 14-19, 2013}}. \bibinfo{publisher}{{IEEE}}, \bibinfo{address}{Turin, Italy}, \bibinfo{pages}{315--319}.
\newblock
\href{https://doi.org/10.1109/INFCOM.2013.6566786}{doi:\nolinkurl{10.1109/INFCOM.2013.6566786}}


\bibitem[Duan et~al\mbox{.}(2019a)]%
        {Duan}
\bibfield{author}{\bibinfo{person}{Ruian Duan}, \bibinfo{person}{Ashish Bijlani}, \bibinfo{person}{Yang Ji}, \bibinfo{person}{Omar Alrawi}, \bibinfo{person}{Yiyuan Xiong}, \bibinfo{person}{Moses Ike}, \bibinfo{person}{Brendan Saltaformaggio}, {and} \bibinfo{person}{Wenke Lee}.} \bibinfo{year}{2019}\natexlab{a}.
\newblock \showarticletitle{Automating Patching of Vulnerable Open-Source Software Versions in Application Binaries}. In \bibinfo{booktitle}{\emph{26th Annual Network and Distributed System Security Symposium, {NDSS} 2019, San Diego, California, USA, February 24-27, 2019}}. \bibinfo{publisher}{The Internet Society}, \bibinfo{address}{San Diego, CA, USA}.
\newblock
\urldef\tempurl%
\url{https://www.ndss-symposium.org/ndss-paper/automating-patching-of-vulnerable-open-source-software-versions-in-application-binaries/}
\showURL{%
\tempurl}


\bibitem[Duan et~al\mbox{.}(2019b)]%
        {YueAG}
\bibfield{author}{\bibinfo{person}{Yue Duan}, \bibinfo{person}{Lian Gao}, \bibinfo{person}{Jie Hu}, {and} \bibinfo{person}{Heng Yin}.} \bibinfo{year}{2019}\natexlab{b}.
\newblock \showarticletitle{Automatic Generation of Non-intrusive Updates for {Third-Party} Libraries in Android Applications}. In \bibinfo{booktitle}{\emph{22nd International Symposium on Research in Attacks, Intrusions and Defenses (RAID 2019)}}. \bibinfo{publisher}{USENIX Association}, \bibinfo{address}{Chaoyang District, Beijing}, \bibinfo{pages}{277--292}.
\newblock
\showISBNx{978-1-939133-07-6}
\urldef\tempurl%
\url{https://www.usenix.org/conference/raid2019/presentation/duan}
\showURL{%
\tempurl}


\bibitem[Espressif(2025a)]%
        {espAB}
\bibfield{author}{\bibinfo{person}{Espressif}.} \bibinfo{year}{2025}\natexlab{a}.
\newblock \bibinfo{title}{{A}/{B} {M}ethod with {O}ver {T}he {A}ir {U}pdates ({O}{T}{A}) in {E}spressif {S}ystems.}
\newblock \bibinfo{howpublished}{\url{https://docs.espressif.com/projects/esp-idf/en/latest/esp32/api-reference/system/ota.html}}.
\newblock


\bibitem[Espressif(2025b)]%
        {espidf}
\bibfield{author}{\bibinfo{person}{Espressif}.} \bibinfo{year}{2025}\natexlab{b}.
\newblock \bibinfo{title}{{E}{S}{P}-{I}{D}{F}.}
\newblock \bibinfo{howpublished}{\url{https://idf.espressif.com/}}.
\newblock


\bibitem[Falliere et~al\mbox{.}(2011)]%
        {Falliere}
\bibfield{author}{\bibinfo{person}{Nicolas Falliere}, \bibinfo{person}{Liam~O Murchu}, \bibinfo{person}{Eric Chien}, {et~al\mbox{.}}} \bibinfo{year}{2011}\natexlab{}.
\newblock \showarticletitle{W32. stuxnet dossier}.
\newblock \bibinfo{journal}{\emph{White paper, symantec corp., security response}} \bibinfo{volume}{5}, \bibinfo{number}{6} (\bibinfo{year}{2011}), \bibinfo{pages}{29}.
\newblock


\bibitem[GigaDevice(2025)]%
        {gd32}
\bibfield{author}{\bibinfo{person}{GigaDevice}.} \bibinfo{year}{2025}\natexlab{}.
\newblock \bibinfo{title}{{G}{D}32{V}{F}103 has 4 harfware breakpoints. 45-46.}
\newblock \bibinfo{howpublished}{\url{https://www.gigadevice.com.cn/Public/Uploads/uploadfile/files/20240403/GD32VF103_Datasheet_Rev1.9.pdf}}.
\newblock


\bibitem[Gu et~al\mbox{.}(2012)]%
        {Tianxiao}
\bibfield{author}{\bibinfo{person}{Tianxiao Gu}, \bibinfo{person}{Chun Cao}, \bibinfo{person}{Chang Xu}, \bibinfo{person}{Xiaoxing Ma}, \bibinfo{person}{Linghao Zhang}, {and} \bibinfo{person}{Jian Lu}.} \bibinfo{year}{2012}\natexlab{}.
\newblock \showarticletitle{Javelus: A Low Disruptive Approach to Dynamic Software Updates}. In \bibinfo{booktitle}{\emph{Proceedings of the 2012 19th Asia-Pacific Software Engineering Conference - Volume 01}} \emph{(\bibinfo{series}{APSEC '12})}. \bibinfo{publisher}{IEEE Computer Society}, \bibinfo{address}{USA}, \bibinfo{pages}{527–536}.
\newblock
\showISBNx{9780769549224}
\href{https://doi.org/10.1109/APSEC.2012.55}{doi:\nolinkurl{10.1109/APSEC.2012.55}}


\bibitem[Halperin et~al\mbox{.}(2008)]%
        {Halperin}
\bibfield{author}{\bibinfo{person}{Daniel Halperin}, \bibinfo{person}{Thomas~S. Heydt{-}Benjamin}, \bibinfo{person}{Benjamin Ransford}, \bibinfo{person}{Shane~S. Clark}, \bibinfo{person}{Benessa Defend}, \bibinfo{person}{Will Morgan}, \bibinfo{person}{Kevin Fu}, \bibinfo{person}{Tadayoshi Kohno}, {and} \bibinfo{person}{William~H. Maisel}.} \bibinfo{year}{2008}\natexlab{}.
\newblock \showarticletitle{Pacemakers and Implantable Cardiac Defibrillators: Software Radio Attacks and Zero-Power Defenses}. In \bibinfo{booktitle}{\emph{2008 {IEEE} Symposium on Security and Privacy {(SP} 2008), 18-21 May 2008, Oakland, California, {USA}}}. \bibinfo{publisher}{{IEEE} Computer Society}, \bibinfo{address}{USA}, \bibinfo{pages}{129--142}.
\newblock
\href{https://doi.org/10.1109/SP.2008.31}{doi:\nolinkurl{10.1109/SP.2008.31}}


\bibitem[Hat(2025)]%
        {Kpatch}
\bibfield{author}{\bibinfo{person}{Red Hat}.} \bibinfo{year}{2025}\natexlab{}.
\newblock \bibinfo{title}{{I}ntroducing kpatch: {D}ynamic {K}ernel {P}atching.}
\newblock \bibinfo{howpublished}{\url{https://www.redhat.com/de/blog/introducing-kpatch-dynamic-kernel-patching}}.
\newblock


\bibitem[Hayden et~al\mbox{.}(2014)]%
        {Hayden}
\bibfield{author}{\bibinfo{person}{Christopher~M. Hayden}, \bibinfo{person}{Karla Saur}, \bibinfo{person}{Edward~K. Smith}, \bibinfo{person}{Michael Hicks}, {and} \bibinfo{person}{Jeffrey~S. Foster}.} \bibinfo{year}{2014}\natexlab{}.
\newblock \showarticletitle{Kitsune: Efficient, General-Purpose Dynamic Software Updating for C}.
\newblock \bibinfo{journal}{\emph{ACM Trans. Program. Lang. Syst.}} \bibinfo{volume}{36}, \bibinfo{number}{4}, Article \bibinfo{articleno}{13} (\bibinfo{date}{Oct.} \bibinfo{year}{2014}), \bibinfo{numpages}{38}~pages.
\newblock
\showISSN{0164-0925}
\href{https://doi.org/10.1145/2629460}{doi:\nolinkurl{10.1145/2629460}}


\bibitem[Hayden et~al\mbox{.}(2012)]%
        {Hayden2012EvaluatingDS}
\bibfield{author}{\bibinfo{person}{Christopher~M. Hayden}, \bibinfo{person}{Edward~K. Smith}, \bibinfo{person}{Eric~A. Hardisty}, \bibinfo{person}{Michael~W. Hicks}, {and} \bibinfo{person}{Jeffrey~S. Foster}.} \bibinfo{year}{2012}\natexlab{}.
\newblock \showarticletitle{Evaluating Dynamic Software Update Safety Using Systematic Testing}.
\newblock \bibinfo{journal}{\emph{IEEE Transactions on Software Engineering}} \bibinfo{volume}{38}, \bibinfo{number}{6} (\bibinfo{year}{2012}), \bibinfo{pages}{1340--1354}.
\newblock


\bibitem[He et~al\mbox{.}(2022)]%
        {YiHe}
\bibfield{author}{\bibinfo{person}{Yi He}, \bibinfo{person}{Zhenhua Zou}, \bibinfo{person}{Kun Sun}, \bibinfo{person}{Zhuotao Liu}, \bibinfo{person}{Ke Xu}, \bibinfo{person}{Qian Wang}, \bibinfo{person}{Chao Shen}, \bibinfo{person}{Zhi Wang}, {and} \bibinfo{person}{Qi Li}.} \bibinfo{year}{2022}\natexlab{}.
\newblock \showarticletitle{{RapidPatch}: Firmware Hotpatching for {Real-Time} Embedded Devices}. In \bibinfo{booktitle}{\emph{31st USENIX Security Symposium (USENIX Security 22)}}. \bibinfo{publisher}{USENIX Association}, \bibinfo{address}{Boston, MA}, \bibinfo{pages}{2225--2242}.
\newblock
\showISBNx{978-1-939133-31-1}
\urldef\tempurl%
\url{https://www.usenix.org/conference/usenixsecurity22/presentation/he-yi}
\showURL{%
\tempurl}


\bibitem[Holmbacka et~al\mbox{.}(2013)]%
        {holmbacka}
\bibfield{author}{\bibinfo{person}{Simon Holmbacka}, \bibinfo{person}{Wictor Lund}, \bibinfo{person}{S{\'{e}}bastien Lafond}, {and} \bibinfo{person}{Johan Lilius}.} \bibinfo{year}{2013}\natexlab{}.
\newblock \showarticletitle{Lightweight Framework for Runtime Updating of C-Based Software in Embedded Systems}. In \bibinfo{booktitle}{\emph{5th Workshop on Hot Topics in Software Upgrades, HotSWUp'13, San Jose, CA, USA, June 28, 2013}}, \bibfield{editor}{\bibinfo{person}{Cristian Cadar} {and} \bibinfo{person}{Jeff Foster}} (Eds.). \bibinfo{publisher}{{USENIX} Association}, \bibinfo{address}{San Jose, CA, USA}.
\newblock
\urldef\tempurl%
\url{https://www.usenix.org/conference/hotswup13/workshop-program/presentation/holmbacka}
\showURL{%
\tempurl}


\bibitem[JSOF(2025)]%
        {Ripple20}
\bibfield{author}{\bibinfo{person}{JSOF}.} \bibinfo{year}{2025}\natexlab{}.
\newblock \bibinfo{title}{{R}ipple20: 19 {Z}ero-{D}ay {V}ulnerabilities {A}mplified by the {S}upply {C}hain}.
\newblock \bibinfo{howpublished}{\url{https://www.jsof-tech.com/ripple20/}}.
\newblock


\bibitem[Labs(2020)]%
        {AMNESIA33}
\bibfield{author}{\bibinfo{person}{Foresout~Research Labs}.} \bibinfo{year}{2020}\natexlab{}.
\newblock \bibinfo{title}{AMNESIA:33 – {F}orescout {R}esearch {L}abs discovered 33 vulnerabilities impacting millions of {I}o{T}, {O}{T} and {I}{T} devices}.
\newblock \bibinfo{howpublished}{\url{https://www.forescout.com/resources/amnesia33-research-report-executive-summary/}}.
\newblock


\bibitem[Li et~al\mbox{.}(2021)]%
        {DEEPRL4FL}
\bibfield{author}{\bibinfo{person}{Yi Li}, \bibinfo{person}{Shaohua Wang}, {and} \bibinfo{person}{Tien~N. Nguyen}.} \bibinfo{year}{2021}\natexlab{}.
\newblock \showarticletitle{Fault Localization with Code Coverage Representation Learning}. In \bibinfo{booktitle}{\emph{Proceedings of the 43rd International Conference on Software Engineering}} (Madrid, Spain) \emph{(\bibinfo{series}{ICSE '21})}. \bibinfo{publisher}{IEEE Press}, \bibinfo{address}{Virtual}, \bibinfo{pages}{661–673}.
\newblock
\showISBNx{9781450390859}
\href{https://doi.org/10.1109/ICSE43902.2021.00067}{doi:\nolinkurl{10.1109/ICSE43902.2021.00067}}


\bibitem[Lou et~al\mbox{.}(2021)]%
        {GRACE}
\bibfield{author}{\bibinfo{person}{Yiling Lou}, \bibinfo{person}{Qihao Zhu}, \bibinfo{person}{Jinhao Dong}, \bibinfo{person}{Xia Li}, \bibinfo{person}{Zeyu Sun}, \bibinfo{person}{Dan Hao}, \bibinfo{person}{Lu Zhang}, {and} \bibinfo{person}{Lingming Zhang}.} \bibinfo{year}{2021}\natexlab{}.
\newblock \showarticletitle{Boosting coverage-based fault localization via graph-based representation learning}. In \bibinfo{booktitle}{\emph{Proceedings of the 29th ACM Joint Meeting on European Software Engineering Conference and Symposium on the Foundations of Software Engineering}} (Athens, Greece) \emph{(\bibinfo{series}{ESEC/FSE 2021})}. \bibinfo{publisher}{Association for Computing Machinery}, \bibinfo{address}{New York, NY, USA}, \bibinfo{pages}{664–676}.
\newblock
\showISBNx{9781450385626}
\href{https://doi.org/10.1145/3468264.3468580}{doi:\nolinkurl{10.1145/3468264.3468580}}


\bibitem[Makris and Bazzi(2009)]%
        {Makris}
\bibfield{author}{\bibinfo{person}{Kristis Makris} {and} \bibinfo{person}{Rida~A. Bazzi}.} \bibinfo{year}{2009}\natexlab{}.
\newblock \showarticletitle{Immediate multi-threaded dynamic software updates using stack reconstruction}. In \bibinfo{booktitle}{\emph{Proceedings of the 2009 Conference on USENIX Annual Technical Conference}} (San Diego, California) \emph{(\bibinfo{series}{USENIX'09})}. \bibinfo{publisher}{USENIX Association}, \bibinfo{address}{USA}, \bibinfo{pages}{31}.
\newblock


\bibitem[Makris and Ryu(2007)]%
        {MakrisD}
\bibfield{author}{\bibinfo{person}{Kristis Makris} {and} \bibinfo{person}{Kyung~Dong Ryu}.} \bibinfo{year}{2007}\natexlab{}.
\newblock \showarticletitle{Dynamic and adaptive updates of non-quiescent subsystems in commodity operating system kernels}. In \bibinfo{booktitle}{\emph{Proceedings of the 2nd ACM SIGOPS/EuroSys European Conference on Computer Systems 2007}} (Lisbon, Portugal) \emph{(\bibinfo{series}{EuroSys '07})}. \bibinfo{publisher}{Association for Computing Machinery}, \bibinfo{address}{New York, NY, USA}, \bibinfo{pages}{327–340}.
\newblock
\showISBNx{9781595936363}
\href{https://doi.org/10.1145/1272996.1273031}{doi:\nolinkurl{10.1145/1272996.1273031}}


\bibitem[MbedTLS(2025)]%
        {MbedTLS}
\bibfield{author}{\bibinfo{person}{MbedTLS}.} \bibinfo{year}{2025}\natexlab{}.
\newblock \bibinfo{title}{{M}bed {T}{L}{S}}.
\newblock \bibinfo{howpublished}{\url{https://github.com/Mbed-TLS/mbedtls}}.
\newblock


\bibitem[Meng et~al\mbox{.}(2022)]%
        {TRANSFER}
\bibfield{author}{\bibinfo{person}{Xiangxin Meng}, \bibinfo{person}{Xu Wang}, \bibinfo{person}{Hongyu Zhang}, \bibinfo{person}{Hailong Sun}, {and} \bibinfo{person}{Xudong Liu}.} \bibinfo{year}{2022}\natexlab{}.
\newblock \showarticletitle{Improving fault localization and program repair with deep semantic features and transferred knowledge}. In \bibinfo{booktitle}{\emph{Proceedings of the 44th International Conference on Software Engineering}} (Pittsburgh, Pennsylvania) \emph{(\bibinfo{series}{ICSE '22})}. \bibinfo{publisher}{Association for Computing Machinery}, \bibinfo{address}{New York, NY, USA}, \bibinfo{pages}{1169–1180}.
\newblock
\showISBNx{9781450392211}
\href{https://doi.org/10.1145/3510003.3510147}{doi:\nolinkurl{10.1145/3510003.3510147}}


\bibitem[MITRE(2025)]%
        {cveWebsite}
\bibfield{author}{\bibinfo{person}{MITRE}.} \bibinfo{year}{2025}\natexlab{}.
\newblock \bibinfo{title}{{C}{V}{E} {W}ebsite}.
\newblock \bibinfo{howpublished}{\url{https://www.cve.org/}}.
\newblock


\bibitem[Necula et~al\mbox{.}(2002)]%
        {necula2002cil}
\bibfield{author}{\bibinfo{person}{George~C. Necula}, \bibinfo{person}{Scott McPeak}, \bibinfo{person}{Shree~Prakash Rahul}, {and} \bibinfo{person}{Westley Weimer}.} \bibinfo{year}{2002}\natexlab{}.
\newblock \showarticletitle{CIL: Intermediate Language and Tools for Analysis and Transformation of C Programs}. In \bibinfo{booktitle}{\emph{Proceedings of the 11th International Conference on Compiler Construction}} \emph{(\bibinfo{series}{CC '02})}. \bibinfo{publisher}{Springer-Verlag}, \bibinfo{address}{Berlin, Heidelberg}, \bibinfo{pages}{213–228}.
\newblock
\showISBNx{3540433694}


\bibitem[Niesler et~al\mbox{.}(2021)]%
        {Niesler}
\bibfield{author}{\bibinfo{person}{Christian Niesler}, \bibinfo{person}{Sebastian Surminski}, {and} \bibinfo{person}{Lucas Davi}.} \bibinfo{year}{2021}\natexlab{}.
\newblock \showarticletitle{HERA: Hotpatching of Embedded Real-time Applications.}. In \bibinfo{booktitle}{\emph{NDSS}}. \bibinfo{publisher}{The Internet Society}, \bibinfo{address}{Virtual}.
\newblock


\bibitem[NVD(2018)]%
        {CVE201816601}
\bibfield{author}{\bibinfo{person}{NVD}.} \bibinfo{year}{2018}\natexlab{}.
\newblock \bibinfo{title}{{C}{V}{E}-2018-16601}.
\newblock \bibinfo{howpublished}{\url{https://nvd.nist.gov/vuln/detail/CVE-2018-16601}}.
\newblock


\bibitem[Panta et~al\mbox{.}(2009)]%
        {Panta}
\bibfield{author}{\bibinfo{person}{Rajesh~Krishna Panta}, \bibinfo{person}{Saurabh Bagchi}, {and} \bibinfo{person}{Samuel~P. Midkiff}.} \bibinfo{year}{2009}\natexlab{}.
\newblock \showarticletitle{Zephyr: efficient incremental reprogramming of sensor nodes using function call indirections and difference computation}. In \bibinfo{booktitle}{\emph{Proceedings of the 2009 Conference on USENIX Annual Technical Conference}} (San Diego, California) \emph{(\bibinfo{series}{USENIX'09})}. \bibinfo{publisher}{USENIX Association}, \bibinfo{address}{USA}, \bibinfo{pages}{32}.
\newblock


\bibitem[Rajput et~al\mbox{.}(2023)]%
        {Rajput}
\bibfield{author}{\bibinfo{person}{Prashant Hari~Narayan Rajput}, \bibinfo{person}{Constantine Doumanidis}, {and} \bibinfo{person}{Michail Maniatakos}.} \bibinfo{year}{2023}\natexlab{}.
\newblock \showarticletitle{ICSPatch: automated vulnerability localization and non-intrusive hotpatching in industrial control systems using data dependence graphs}. In \bibinfo{booktitle}{\emph{Proceedings of the 32nd USENIX Conference on Security Symposium}} (Anaheim, CA, USA) \emph{(\bibinfo{series}{SEC '23})}. \bibinfo{publisher}{USENIX Association}, \bibinfo{address}{USA}, Article \bibinfo{articleno}{384}, \bibinfo{numpages}{16}~pages.
\newblock
\showISBNx{978-1-939133-37-3}


\bibitem[Ramaswamy et~al\mbox{.}(2010)]%
        {Ramaswamy}
\bibfield{author}{\bibinfo{person}{Ashwin Ramaswamy}, \bibinfo{person}{Sergey Bratus}, \bibinfo{person}{Sean~W. Smith}, {and} \bibinfo{person}{Michael~E. Locasto}.} \bibinfo{year}{2010}\natexlab{}.
\newblock \showarticletitle{Katana: {A} Hot Patching Framework for {ELF} Executables}. In \bibinfo{booktitle}{\emph{{ARES} 2010, Fifth International Conference on Availability, Reliability and Security, 15-18 February 2010, Krakow, Poland}}. \bibinfo{publisher}{{IEEE} Computer Society}, \bibinfo{address}{Krakow, Poland}, \bibinfo{pages}{507--512}.
\newblock
\href{https://doi.org/10.1109/ARES.2010.112}{doi:\nolinkurl{10.1109/ARES.2010.112}}


\bibitem[Rommel et~al\mbox{.}(2020)]%
        {Florian}
\bibfield{author}{\bibinfo{person}{Florian Rommel}, \bibinfo{person}{Christian Dietrich}, \bibinfo{person}{Birte Friesel}, \bibinfo{person}{Marcel K{\"o}ppen}, \bibinfo{person}{Christoph Borchert}, \bibinfo{person}{Michael M{\"u}ller}, \bibinfo{person}{Olaf Spinczyk}, {and} \bibinfo{person}{Daniel Lohmann}.} \bibinfo{year}{2020}\natexlab{}.
\newblock \showarticletitle{From Global to Local Quiescence: {Wait-Free} Code Patching of {Multi-Threaded} Processes}. In \bibinfo{booktitle}{\emph{14th USENIX Symposium on Operating Systems Design and Implementation (OSDI 20)}}. \bibinfo{publisher}{USENIX Association}, \bibinfo{address}{USA}, \bibinfo{pages}{651--666}.
\newblock
\showISBNx{978-1-939133-19-9}
\urldef\tempurl%
\url{https://www.usenix.org/conference/osdi20/presentation/rommel}
\showURL{%
\tempurl}


\bibitem[Salehi and Pattabiraman(2024)]%
        {Salehi}
\bibfield{author}{\bibinfo{person}{Mohsen Salehi} {and} \bibinfo{person}{Karthik Pattabiraman}.} \bibinfo{year}{2024}\natexlab{}.
\newblock \showarticletitle{AutoPatch: Automated Generation of Hotpatches for Real-Time Embedded Devices}. In \bibinfo{booktitle}{\emph{Proceedings of the 2024 on ACM SIGSAC Conference on Computer and Communications Security}} (Salt Lake City, UT, USA) \emph{(\bibinfo{series}{CCS '24})}. \bibinfo{publisher}{Association for Computing Machinery}, \bibinfo{address}{New York, NY, USA}, \bibinfo{pages}{2370–2384}.
\newblock
\showISBNx{9798400706363}
\href{https://doi.org/10.1145/3658644.3690255}{doi:\nolinkurl{10.1145/3658644.3690255}}


\bibitem[Services(2025a)]%
        {freertos}
\bibfield{author}{\bibinfo{person}{Amazon~Web Services}.} \bibinfo{year}{2025}\natexlab{a}.
\newblock \bibinfo{title}{{A}mazon {F}ree{R}{T}{O}{S}}.
\newblock \bibinfo{howpublished}{\url{https://aws.amazon.com/freertos/}}.
\newblock


\bibitem[Services(2025b)]%
        {freertosOTA}
\bibfield{author}{\bibinfo{person}{Amazon~Web Services}.} \bibinfo{year}{2025}\natexlab{b}.
\newblock \bibinfo{title}{{A}{W}{S} {I}o{T} {O}ver the {A}ir ({O}{T}{A}).}
\newblock \bibinfo{howpublished}{\url{https://www.freertos.org/ota/index.html}}.
\newblock


\bibitem[Services(2025c)]%
        {freertosIdle}
\bibfield{author}{\bibinfo{person}{Amazon~Web Services}.} \bibinfo{year}{2025}\natexlab{c}.
\newblock \bibinfo{title}{{F}ree{R}{T}{O}{S} idle task.}
\newblock \bibinfo{howpublished}{\url{https://www.freertos.org/Documentation/02-Kernel/02-Kernel-features/01-Tasks-and-co-routines/15-Idle-task}}.
\newblock


\bibitem[STMicroelectronics(2025)]%
        {STM32CubeFunctionPack}
\bibfield{author}{\bibinfo{person}{STMicroelectronics}.} \bibinfo{year}{2025}\natexlab{}.
\newblock \bibinfo{title}{{S}{T}{M}32{C}ube{F}unction{P}ack\_{P}{L}{C}{W}{I}{F}{I}}.
\newblock \bibinfo{howpublished}{\url{https://github.com/yisea123/STM32CubeFunctionPack_PLCWIFI1_V1.0.1/tree/master}}.
\newblock


\bibitem[SUSE(2025)]%
        {kgraft}
\bibfield{author}{\bibinfo{person}{SUSE}.} \bibinfo{year}{2025}\natexlab{}.
\newblock \bibinfo{title}{Live patching the Linux kernel using kGraft.}
\newblock \bibinfo{howpublished}{\url{https://documentation.suse.com/sles/12-SP5/html/SLES-all/cha-kgraft.html}}.
\newblock


\bibitem[Wahler and Oriol(2014)]%
        {wahlerDF}
\bibfield{author}{\bibinfo{person}{Michael Wahler} {and} \bibinfo{person}{Manuel Oriol}.} \bibinfo{year}{2014}\natexlab{}.
\newblock \showarticletitle{Disruption-free software updates in automation systems}. In \bibinfo{booktitle}{\emph{Proceedings of the 2014 {IEEE} Emerging Technology and Factory Automation, {ETFA} 2014, Barcelona, Spain, September 16-19, 2014}}, \bibfield{editor}{\bibinfo{person}{Antoni Grau} {and} \bibinfo{person}{Herminio Mart{\'{\i}}nez}} (Eds.). \bibinfo{publisher}{{IEEE}}, \bibinfo{address}{Barcelona, Spain}, \bibinfo{pages}{1--8}.
\newblock
\href{https://doi.org/10.1109/ETFA.2014.7005075}{doi:\nolinkurl{10.1109/ETFA.2014.7005075}}


\bibitem[Wahler et~al\mbox{.}(2011)]%
        {wahlerNL}
\bibfield{author}{\bibinfo{person}{Michael Wahler}, \bibinfo{person}{Stefan Richter}, \bibinfo{person}{Sumit Kumar}, {and} \bibinfo{person}{Manuel Oriol}.} \bibinfo{year}{2011}\natexlab{}.
\newblock \showarticletitle{Non-disruptive large-scale component updates for real-time controllers}. In \bibinfo{booktitle}{\emph{Workshops Proceedings of the 27th International Conference on Data Engineering, {ICDE} 2011, April 11-16, 2011, Hannover, Germany}}, \bibfield{editor}{\bibinfo{person}{Serge Abiteboul}, \bibinfo{person}{Klemens B{\"{o}}hm}, \bibinfo{person}{Christoph Koch}, {and} \bibinfo{person}{Kian{-}Lee Tan}} (Eds.). \bibinfo{publisher}{{IEEE} Computer Society}, \bibinfo{address}{Hannover, Germany}, \bibinfo{pages}{174--178}.
\newblock
\href{https://doi.org/10.1109/ICDEW.2011.5767631}{doi:\nolinkurl{10.1109/ICDEW.2011.5767631}}


\bibitem[Wahler et~al\mbox{.}(2009)]%
        {WahlerDS}
\bibfield{author}{\bibinfo{person}{Michael Wahler}, \bibinfo{person}{Stefan Richter}, {and} \bibinfo{person}{Manuel Oriol}.} \bibinfo{year}{2009}\natexlab{}.
\newblock \showarticletitle{Dynamic software updates for real-time systems}. In \bibinfo{booktitle}{\emph{Proceedings of the 2nd International Workshop on Hot Topics in Software Upgrades}} (Orlando, Florida) \emph{(\bibinfo{series}{HotSWUp '09})}. \bibinfo{publisher}{Association for Computing Machinery}, \bibinfo{address}{New York, NY, USA}, Article \bibinfo{articleno}{2}, \bibinfo{numpages}{6}~pages.
\newblock
\showISBNx{9781605587233}
\href{https://doi.org/10.1145/1656437.1656440}{doi:\nolinkurl{10.1145/1656437.1656440}}


\bibitem[WolfSSL(2025)]%
        {wolfSSL}
\bibfield{author}{\bibinfo{person}{WolfSSL}.} \bibinfo{year}{2025}\natexlab{}.
\newblock \bibinfo{title}{{W}olf{S}{S}{L} {E}mbedded {S}{S}{L}/{T}{L}{S} {L}ibrary}.
\newblock \bibinfo{howpublished}{\url{https://www.wolfssl.com/}}.
\newblock


\bibitem[Xu et~al\mbox{.}(2020)]%
        {Zhengzi}
\bibfield{author}{\bibinfo{person}{Zhengzi Xu}, \bibinfo{person}{Yulong Zhang}, \bibinfo{person}{Longri Zheng}, \bibinfo{person}{Liangzhao Xia}, \bibinfo{person}{Chenfu Bao}, \bibinfo{person}{Zhi Wang}, {and} \bibinfo{person}{Yang Liu}.} \bibinfo{year}{2020}\natexlab{}.
\newblock \showarticletitle{Automatic Hot Patch Generation for Android Kernels}. In \bibinfo{booktitle}{\emph{29th USENIX Security Symposium (USENIX Security 20)}}. \bibinfo{publisher}{USENIX Association}, \bibinfo{address}{Boston, MA, USA}, \bibinfo{pages}{2397--2414}.
\newblock
\showISBNx{978-1-939133-17-5}
\urldef\tempurl%
\url{https://www.usenix.org/conference/usenixsecurity20/presentation/xu}
\showURL{%
\tempurl}


\bibitem[Yang et~al\mbox{.}(2024)]%
        {Yang2024LLMfTFL}
\bibfield{author}{\bibinfo{person}{Aidan Z.~H. Yang}, \bibinfo{person}{Claire Le~Goues}, \bibinfo{person}{Ruben Martins}, {and} \bibinfo{person}{Vincent Hellendoorn}.} \bibinfo{year}{2024}\natexlab{}.
\newblock \showarticletitle{Large Language Models for Test-Free Fault Localization}. In \bibinfo{booktitle}{\emph{Proceedings of the IEEE/ACM 46th International Conference on Software Engineering}} (Lisbon, Portugal) \emph{(\bibinfo{series}{ICSE '24})}. \bibinfo{publisher}{Association for Computing Machinery}, \bibinfo{address}{New York, NY, USA}, Article \bibinfo{articleno}{17}, \bibinfo{numpages}{12}~pages.
\newblock
\showISBNx{9798400702174}
\href{https://doi.org/10.1145/3597503.3623342}{doi:\nolinkurl{10.1145/3597503.3623342}}


\bibitem[Ye et~al\mbox{.}(2024)]%
        {Zhenyu}
\bibfield{author}{\bibinfo{person}{Zhenyu Ye}, \bibinfo{person}{Lei Zhou}, \bibinfo{person}{Fengwei Zhang}, \bibinfo{person}{Wenqiang Jin}, \bibinfo{person}{Zhenyu Ning}, \bibinfo{person}{Yupeng Hu}, {and} \bibinfo{person}{Zheng Qin}.} \bibinfo{year}{2024}\natexlab{}.
\newblock \showarticletitle{FortifyPatch: Towards Tamper-Resistant Live Patching in Linux-Based Hypervisor}. In \bibinfo{booktitle}{\emph{Proceedings of the 33rd ACM SIGSOFT International Symposium on Software Testing and Analysis}} (Vienna, Austria) \emph{(\bibinfo{series}{ISSTA 2024})}. \bibinfo{publisher}{Association for Computing Machinery}, \bibinfo{address}{New York, NY, USA}, \bibinfo{pages}{38–50}.
\newblock
\showISBNx{9798400706127}
\href{https://doi.org/10.1145/3650212.3652108}{doi:\nolinkurl{10.1145/3650212.3652108}}


\bibitem[Zephyr(2025a)]%
        {zephyrota}
\bibfield{author}{\bibinfo{person}{Zephyr}.} \bibinfo{year}{2025}\natexlab{a}.
\newblock \bibinfo{title}{{O}ver-the-{A}ir {U}pdate in {Z}ephyr {O}{S}}.
\newblock \bibinfo{howpublished}{\url{https://docs.zephyrproject.org/latest/services/device_mgmt/ota.html}}.
\newblock


\bibitem[Zephyr(2025b)]%
        {zephyr}
\bibfield{author}{\bibinfo{person}{Zephyr}.} \bibinfo{year}{2025}\natexlab{b}.
\newblock \bibinfo{title}{{Z}ephyr {O}{S}}.
\newblock \bibinfo{howpublished}{\url{https://www.zephyrproject.org/}}.
\newblock


\bibitem[Zhang et~al\mbox{.}(2017)]%
        {Xuewen}
\bibfield{author}{\bibinfo{person}{Xuewen Zhang}, \bibinfo{person}{Yuanyuan Zhang}, \bibinfo{person}{Juanru Li}, \bibinfo{person}{Yikun Hu}, \bibinfo{person}{Huayi Li}, {and} \bibinfo{person}{Dawu Gu}.} \bibinfo{year}{2017}\natexlab{}.
\newblock \showarticletitle{Embroidery: Patching Vulnerable Binary Code of Fragmentized Android Devices}. In \bibinfo{booktitle}{\emph{2017 IEEE International Conference on Software Maintenance and Evolution (ICSME)}}. \bibinfo{publisher}{{IEEE} Computer Society}, \bibinfo{address}{Shanghai, China}, \bibinfo{pages}{47--57}.
\newblock
\href{https://doi.org/10.1109/ICSME.2017.15}{doi:\nolinkurl{10.1109/ICSME.2017.15}}


\bibitem[Zhou et~al\mbox{.}(2024)]%
        {MingZhou}
\bibfield{author}{\bibinfo{person}{Ming Zhou}, \bibinfo{person}{Haining Wang}, \bibinfo{person}{Ke Li}, \bibinfo{person}{Hongsong Zhu}, {and} \bibinfo{person}{Limin Sun}.} \bibinfo{year}{2024}\natexlab{}.
\newblock \showarticletitle{Save the Bruised Striver: A Reliable Live Patching Framework for Protecting Real-World PLCs}. In \bibinfo{booktitle}{\emph{Proceedings of the Nineteenth European Conference on Computer Systems}} (Athens, Greece) \emph{(\bibinfo{series}{EuroSys '24})}. \bibinfo{publisher}{Association for Computing Machinery}, \bibinfo{address}{New York, NY, USA}, \bibinfo{pages}{1192–1207}.
\newblock
\showISBNx{9798400704376}
\href{https://doi.org/10.1145/3627703.3650068}{doi:\nolinkurl{10.1145/3627703.3650068}}


\bibitem[Zurawski(2018)]%
        {Zurawski}
\bibfield{author}{\bibinfo{person}{Richard Zurawski}.} \bibinfo{year}{2018}\natexlab{}.
\newblock \bibinfo{booktitle}{\emph{Embedded Systems Handbook.}}
\newblock \bibinfo{publisher}{CRC press}, \bibinfo{address}{Boca Raton}.
\newblock


\bibitem[Zynamics(2022)]%
        {Bindiff}
\bibfield{author}{\bibinfo{person}{Zynamics}.} \bibinfo{year}{2022}\natexlab{}.
\newblock \bibinfo{title}{{B}in{D}iff {H}omepage.}
\newblock \bibinfo{howpublished}{\url{https://www.zynamics.com/}}.
\newblock


\end{thebibliography}


\appendix
\thispagestyle{empty}


\section{Appendix}
\label{section:appendix}
\subsection{Repair effectiveness on the RTOS vulnerabilities}
\label{appendix:repair_effectiveness}
Table~\ref{tab:CVEtable} shows the StackPatch's effectiveness for repairing RTOS vulnerabilities on three embedded architectures. 
We studied 107 vulnerabilities on two real-time operating systems (FreeRTOS and ZephyrOS) and five embedded libraries (PicoTCP, mbedTLS, uIP, wolfSSL, and Contiki), covering a total of 30 software versions. 
The study identified 88 high-risk vulnerabilities (CVSS v3 scores 7.0-10.0), with vulnerability data sorted in ascending order by CVE-ID. 
The experiments were conducted on three embedded architecture platforms: two ARM-based development boards (STM32F401 and nRF52840), an RISC-V-based GD32VF103 development board, and an Xtensa-based ESP32S3 development board. 
The types of vulnerability patches were categorized into three types: local variables (L), global variables (G), and macro variables (M), providing an important dimension for understanding the patching patterns of embedded system security vulnerabilities. 
Note that a "-" in the embedded architecture column indicates the vulnerability does not exist for that architecture, while a "-" in the patch type column signifies that the generated hot patch cannot fix the vulnerability. 

Among the 107 vulnerabilities, StackPatch repaired 102, while five could not be addressed, the repair success rate was 95.3\%. 

\begin{table*}[t] 
    \footnotesize
    \centering
     \setlength{\tabcolsep}{1.35mm}{
    \begin{tabular}{|l|l|l|c|c|c|c|c||l|l|l|c|c|c|c|c|}
             \hline
\multirow{2}{*}{CVE-ID} & \multirow{2}{*}{OS/Lib} & \multirow{2}{*}{Version} & \multicolumn{1}{c|}{\multirow{2}{*}{CVSS}} & \multicolumn{3}{c|}{Embedded architecture}     & \multirow{2}{*}{\makecell[c]{Patch \\ type}} & \multirow{2}{*}{CVE-ID} & \multirow{2}{*}{OS/Lib} & \multirow{2}{*}{Version} & \multicolumn{1}{c|}{\multirow{2}{*}{CVSS}} & \multicolumn{3}{c|}{Embedded architecture}     & \multirow{2}{*}{\makecell[c]{Patch \\ type}} \\\cline{5-7} \cline{13-15}
                        &                         &                                     &                      & ARM & RISC-V  & Xtensa  &                             &                         &                         &                                     &                       & ARM   & RISC-V  & Xtensa  &                             \\\hline
CVE-2018-16522          & FreeRTOS                & V1.3.1                              & 8.1                                       & \Checkmark&\Checkmark&\Checkmark & L& CVE-2021-3581           & ZephyrOS                & V2.5.0                              & 8.8                                       & \Checkmark&\Checkmark&\Checkmark & L                           \\\hline
CVE-2018-16523          & FreeRTOS                & V10.0.1                             & 7.4                                       & \Checkmark&\Checkmark&\Checkmark &L& CVE-2021-3861           & ZephyrOS                & V2.6.0                              & 8.2                                 &\Checkmark&\Checkmark&\Checkmark         & L G                         \\\hline
CVE-2018-16524          & FreeRTOS                & V10.0.1                             & 5.9                                       & \Checkmark&\Checkmark&\Checkmark &L   & CVE-2022-1042           & ZephyrOS                & V3.0.0                              & 8.8                                       & \Checkmark&\Checkmark&\Checkmark & L M                         \\\hline
CVE-2018-16525          & FreeRTOS                & V10.0.1                             & 8.1                                       & \Checkmark&\Checkmark&\Checkmark &L  & CVE-2022-1841           & ZephyrOS                & V3.0.0                              & 7.2                                       & \Checkmark&\Checkmark&\Checkmark & L                           \\\hline
CVE-2018-16526          & FreeRTOS                & V10.0.1                             & 8.1                                       & \Checkmark&\Checkmark&\Checkmark &L     & CVE-2022-2741           & ZephyrOS                & V3.1.0                              & 8.2                                       & \ding{55}&\ding{55}&\ding{55}                         & -                           \\\hline
CVE-2018-16527          & FreeRTOS                & V10.0.1                             & 5.9                                       & \Checkmark&\Checkmark&\Checkmark &  L           & CVE-2022-2993           & ZephyrOS                & V3.1.0                              & 9.8                                       & \Checkmark&\Checkmark&\Checkmark & L                           \\\hline
CVE-2018-16528          & FreeRTOS                & V1.3.1                              & 8.1                                       & \Checkmark&\Checkmark&\Checkmark &   L  & CVE-2023-3725           & ZephyrOS                & V3.4.0                              & 9.8                                       & \Checkmark&\Checkmark&\Checkmark & L                           \\\hline
CVE-2018-16598          & FreeRTOS                & V10.0.1                             & 5.9                                       & \Checkmark&\Checkmark&\Checkmark &   L & CVE-2023-4257           & ZephyrOS                & V3.4.0                              & 9.8                                       & \Checkmark&\Checkmark&\Checkmark & L M                         \\\hline
CVE-2018-16599          & FreeRTOS                & V10.0.1                             & 5.9                                       & \Checkmark&\Checkmark&\Checkmark &L   & CVE-2023-4259           & ZephyrOS                & V3.4.0                              & 8.8                                       & \Checkmark&\Checkmark&\Checkmark & L                           \\\hline
CVE-2018-16600          & FreeRTOS                & V10.0.1                             & 5.9                                       & \Checkmark&\Checkmark&\Checkmark &  L      & CVE-2023-4260           & ZephyrOS                & V3.4.0                              & 10                                        & \Checkmark&\Checkmark&\Checkmark & L                           \\\hline
CVE-2018-16601          & FreeRTOS                & V10.0.1                             & 8.1                                       & \Checkmark&\Checkmark&\Checkmark & L    & CVE-2023-4263           & ZephyrOS                & V3.4.0                              & 8.8                                       & \Checkmark&\Checkmark&\Checkmark & L                           \\\hline
CVE-2018-16602          & FreeRTOS                & V10.0.1                             & 5.9                                       & \Checkmark&\Checkmark&\Checkmark & L & CVE-2023-4265           & ZephyrOS                & V3.4.0                              & 6.8                                       & \Checkmark&\Checkmark&\Checkmark & L                           \\\hline
CVE-2018-16603          & FreeRTOS                & V10.0.1                             & 5.9                                       & \Checkmark&\Checkmark&\Checkmark & L & CVE-2023-4424           & ZephyrOS                & V3.4.0                              & 8.8                                       & \Checkmark&\Checkmark&\Checkmark & L                           \\\hline
CVE-2019-18178          & FreeRTOS                & V10.0.1                             & 7.5                                       & \Checkmark&\Checkmark&\Checkmark & L & CVE-2023-5055           & ZephyrOS                & V3.4.0                              & 9.8                                       & \Checkmark&\Checkmark&\Checkmark & L                           \\\hline
CVE-2019-13120          & FreeRTOS                & V1.4.8                              & 7.5                                       & \Checkmark&\Checkmark&\Checkmark & L & CVE-2023-5139           & ZephyrOS                & V3.4.0                              & 7.8                                       & \Checkmark&\Checkmark&\Checkmark & L                           \\\hline
CVE-2021-31571          & FreeRTOS                & V10.4.3                             & 9.8                                       & \Checkmark&\Checkmark&\Checkmark & L & CVE-2023-5184           & ZephyrOS                & V3.4.0                              & 8.8                                       & \Checkmark&\Checkmark&\Checkmark & L                           \\\hline
CVE-2021-31572          & FreeRTOS                & V10.4.3                             & 9.8                                       & \Checkmark&\Checkmark&\Checkmark & L & CVE-2023-5779           & ZephyrOS                & V3.5.0                              & 9.8                                       & \Checkmark&\Checkmark&\Checkmark & L                           \\\hline
CVE-2021-32020          & FreeRTOS                & V10.4.3                              & 9.8                                       & \Checkmark&\Checkmark&\Checkmark & L & CVE-2023-6749           & ZephyrOS                & V3.5.0                              & 9.8                                       & \Checkmark&\Checkmark&\Checkmark & L                           \\\hline
CVE-2021-42553          & FreeRTOS                & V3.5.1                               & 9.8                                       & \Checkmark&\Checkmark&\Checkmark & L & CVE-2023-6881           & ZephyrOS                & V3.5.0                              & 7.3                                       & \Checkmark&\Checkmark&\Checkmark & L                           \\\hline
CVE-2024-2212           & FreeRTOS                & V6.4.0                              & 7.8                                       & \Checkmark&\Checkmark&\Checkmark & L & CVE-2024-5931           & ZephyrOS                & V3.6.0                              & 6.3                                       & \Checkmark&\Checkmark&\Checkmark & L                           \\\hline
CVE-2024-38373          & FreeRTOS                & V4.0.0                               & 9.6                                       & \Checkmark&\Checkmark&\Checkmark & L & CVE-2024-6135           & ZephyrOS                & V3.6.0                              & 7.6                                       & \Checkmark&\Checkmark&\Checkmark & L                           \\\hline
CVE-2017-14199          & ZephyrOS                & V1.10.0                             & 9.8                                       & \Checkmark&\Checkmark&\Checkmark &  L & CVE-2024-6259           & ZephyrOS                & V3.6.0                              & 7.6                                       & \Checkmark&\Checkmark&\Checkmark & L                           \\\hline
CVE-2017-14201          & ZephyrOS                & V1.14.0                             & 7.8                                       & \Checkmark&\Checkmark&\Checkmark & L & CVE-2024-6443           & ZephyrOS                & V3.6.0                              & 6.3                                       & \Checkmark&\Checkmark&\Checkmark & L                           \\\hline
CVE-2017-14202          & ZephyrOS                & V1.14.0                             & 7.8                                       & \Checkmark&\Checkmark&\Checkmark & L M                         & CVE-2024-6444           & ZephyrOS                & V3.6.0                              & 6.3                                       & \Checkmark&\Checkmark&\Checkmark & L                           \\\hline
CVE-2019-9506           & ZephyrOS                & V5.1.0                              & 8.1                                       & \Checkmark&\Checkmark&\Checkmark & L M & CVE-2024-8798           & ZephyrOS                & V3.7.0                              & 7.5                                       & -& \Checkmark &- & L                           \\\hline
CVE-2020-10019          & ZephyrOS                & V2.1.0                              & 8.1                                       & \Checkmark&\Checkmark&\Checkmark & L & CVE-2024-10395          & ZephyrOS                & V3.7.0                              & 8.6                                       & \Checkmark&\Checkmark&\Checkmark & L                           \\\hline
CVE-2020-10021          & ZephyrOS                & V2.1.0                              & 8.1                                       & \Checkmark&\Checkmark&\Checkmark & L & CVE-2024-11263          & ZephyrOS                & V3.7.0                              & 9.3                                       & \ding{55}&\ding{55}&\ding{55}                         & -                           \\\hline
CVE-2020-10022          & ZephyrOS                & V2.1.0                              & 9.8                                       & \Checkmark&\Checkmark&\Checkmark &  L M G & CVE-2020-17441          & PicoTCP                 & V1.7.0                              & 9.0                                       & \Checkmark&\Checkmark&\Checkmark & L                            \\\hline
CVE-2020-10023          & ZephyrOS                & V2.1.0                              & 6.9                                       & \Checkmark&\Checkmark&\Checkmark &  L& CVE-2020-17442          & PicoTCP                 & V1.7.0                              & 7.5                                       & \Checkmark&\Checkmark&\Checkmark &   L                          \\\hline
CVE-2020-10024          & ZephyrOS                & V2.1.0                              & 7.8                                       & \Checkmark& - & -  & L & CVE-2020-17443          & PicoTCP                 & V1.7.0                              & 7.5                                       & \Checkmark&\Checkmark&\Checkmark &    L                         \\\hline
CVE-2020-10027          & ZephyrOS                & V2.1.0                              & 7.8                                       & \Checkmark & -& - & L & CVE-2020-17444          & PicoTCP                 & V1.7.0                              & 7.5                                       & \Checkmark&\Checkmark&\Checkmark &      L                       \\\hline
CVE-2020-10028          & ZephyrOS                & V2.1.0                              & 7.8                                       & \Checkmark&\Checkmark&\Checkmark & L & CVE-2020-17445          & PicoTCP                 & V1.7.0                              & 7.5                                       & \Checkmark&\Checkmark&\Checkmark &      L                       \\\hline
CVE-2020-10058          & ZephyrOS                & V2.1.0                              & 7.8                                       & \Checkmark&\Checkmark&\Checkmark & L & CVE-2020-24337          & PicoTCP                 & V1.7.0                              & 7.5                                       & \Checkmark&\Checkmark&\Checkmark &        L                     \\\hline
CVE-2020-10059          & ZephyrOS                & V2.1.0                              & 5.8                                       & \Checkmark&\Checkmark&\Checkmark & L & CVE-2020-24338          & PicoTCP                 & V1.7.0                              & 9.8                                       & \Checkmark&\Checkmark&\Checkmark &       L                      \\\hline
CVE-2020-10060          & ZephyrOS                & V2.1.0                              & 8.0                                       & \Checkmark&\Checkmark&\Checkmark & L & CVE-2020-24339          & PicoTCP                 & V1.7.0                              & 7.5                                       & \Checkmark&\Checkmark&\Checkmark &      L                       \\\hline
CVE-2020-10061          & ZephyrOS                & V2.2.0                              & 8.8                                       & \Checkmark&\Checkmark&\Checkmark & L & CVE-2021-33304          & PicoTCP                 & V1.7.0                              & 9.8                                       & \Checkmark&\Checkmark&\Checkmark &      L                        \\\hline
CVE-2020-10062          & ZephyrOS                & V2.2.0                              & 9.8                                       & \Checkmark&\Checkmark&\Checkmark & L G & CVE-2023-30463          & PicoTCP                 & V1.7.0                              & 7.5                                       & \Checkmark&\Checkmark&\Checkmark &      L                        \\\hline
CVE-2020-10063          & ZephyrOS                & V2.2.0                              & 7.5                                       & \Checkmark&\Checkmark&\Checkmark & L & CVE-2023-35846          & PicoTCP                 & V2.1.0                              & 7.5                                       & \Checkmark&\Checkmark&\Checkmark &      L                        \\\hline
CVE-2020-10064          & ZephyrOS                & V2.2.0                              & 9.8                                       &\ding{55}&\ding{55}&\ding{55}                       & -                           & CVE-2023-35847          & PicoTCP                 & V2.1.0                              & 7.5                                       & \Checkmark&\Checkmark&\Checkmark &    L     \\\hline
CVE-2020-10066          & ZephyrOS                & V2.2.0                              & 5.7                                       & \Checkmark&\Checkmark&\Checkmark & L & CVE-2017-2784           & mbedTLS                 & V2.4.2                              & 8.1                                       & \Checkmark&\Checkmark&\Checkmark &    L                          \\\hline
CVE-2020-10067          & ZephyrOS                & V2.1.0                              & 7.8                                       & \Checkmark&\Checkmark&\Checkmark & L & CVE-2020-17437          & uIP                     & V1.0                                & 8.2                                       & \Checkmark&\Checkmark&\Checkmark &          L                    \\\hline
CVE-2020-10068          & ZephyrOS                & V2.2.0                              & 6.5                                       & \Checkmark&\Checkmark&\Checkmark & L & CVE-2020-24334          & uIP                     & V1.0                                & 8.2                                       & \Checkmark&\Checkmark&\Checkmark &      L                        \\\hline
CVE-2020-10069          & ZephyrOS                & V2.2.0                              & 6.5                                       & \Checkmark&\Checkmark&\Checkmark & L & CVE-2020-24335          & uIP                     & V1.0                                & 7.5                                       & \Checkmark&\Checkmark&\Checkmark &      L                        \\\hline
CVE-2020-10070          & ZephyrOS                & V2.2.0                              & 9.8                                       & \Checkmark&\Checkmark&\Checkmark & L & CVE-2019-16748          & wolfSSL                 & V4.1.0                              & 9.8                                       & \Checkmark&\Checkmark&\Checkmark &     L                         \\\hline
CVE-2020-10071          & ZephyrOS                & V2.2.0                              & 9.8                                       & \Checkmark&\Checkmark&\Checkmark &  L G & CVE-2019-18840          & wolfSSL                 & V4.1.0                              & 7.5                                       & \Checkmark&\Checkmark&\Checkmark & L                           \\\hline
CVE-2020-10072          & ZephyrOS                & V2.2.0                              & 5.3                                       & \ding{55}&\ding{55}&\ding{55}                         & -                           & CVE-2020-12457          & wolfSSL                 & V4.5.0                              & 7.5                                       & \Checkmark&\Checkmark&\Checkmark &     L                         \\\hline
CVE-2020-13598          & ZephyrOS                & V2.3.0                              & 7.8                                       & \ding{55}&\ding{55}&\ding{55}                         & -                           & CVE-2020-24585          & wolfSSL                 & V4.5.0                              & 5.3                                       & \Checkmark&\Checkmark&\Checkmark &      L                        \\\hline
CVE-2020-13600          & ZephyrOS                & V2.3.0                              & 7.6                                       & \Checkmark&\Checkmark&\Checkmark & L M & CVE-2021-3336           & wolfSSL                 & V4.7.0                              & 8.1                                       & \Checkmark&\Checkmark&\Checkmark & L                             \\\hline
CVE-2020-13601          & ZephyrOS                & V2.3.0                              & 9.8                                       & \Checkmark&\Checkmark&\Checkmark &  L G& CVE-2024-0901           & wolfSSL                 & V5.6.3                              & 7.5                                       & \Checkmark&\Checkmark&\Checkmark &L                              \\\hline
CVE-2020-13602          & ZephyrOS                & V2.2.0                              & 5.5                                       & \Checkmark&\Checkmark&\Checkmark & L & CVE-2020-13987          & Contiki                 & V3.0                                & 7.5                                       & \Checkmark&\Checkmark&\Checkmark &  L                            \\\hline
CVE-2021-3323           & ZephyrOS                & V2.4.0                              & 9.8                                       & \Checkmark&\Checkmark&\Checkmark & L                           & CVE-2020-17439          & Contiki                 & V3.0                                & 8.3                                       & \Checkmark&\Checkmark&\Checkmark &L                              \\\hline
CVE-2021-3430           & ZephyrOS                & V1.14                               & 7.5                                       & \Checkmark&\Checkmark&\Checkmark & L                           & CVE-2020-24336          & Contiki                 & V3.0                                & 9.8                                       & \Checkmark&\Checkmark&\Checkmark &L                              \\\hline
CVE-2021-3434           & ZephyrOS                & V2.5.0                              & 7.8                                       & \Checkmark&\Checkmark&\Checkmark & L                           & CVE-2020-25111          & Contiki                 & V3.0                                & 9.8                                       & \Checkmark&\Checkmark&\Checkmark & L                             \\\hline
CVE-2021-3454           & ZephyrOS                & V2.4.0                              & 7.5                                       & \Checkmark&\Checkmark&\Checkmark & L                           &                         &                         &                                     &                                           & \multicolumn{3}{c|}{}                          &        \\\hline
    \end{tabular}}
    \caption{The effectiveness of StackPatch in defending against common RTOS vulnerabilities on STM32F401, nRF52840, GD32VF103, and ESP32S3 boards. Of the 107 vulnerabilities, 102 were successfully repaired (\Checkmark), while five could not be addressed (\ding{55}).}
    \label{tab:CVEtable}
    
\end{table*}

\end{sloppypar}
\end{document}